\documentclass[a4paper,11pt]{article}
\pdfoutput=1 

\usepackage{jinstpub}
\usepackage[T1]{fontenc} 
\usepackage{graphicx}
\usepackage{caption}
\usepackage{subcaption}
\usepackage{float}
\usepackage{placeins}
\usepackage{booktabs}
\usepackage{lineno}

\title{\boldmath Artificial Neural Networks-based Track Fitting of Cosmic Muons through Stacked Resistive Plate Chambers}

\author[a,1]{Deepak Samuel,\note{Corresponding author.}}
\author[a]{Karthik Suresh,}

\affiliation[a]{Central University of Karnataka,\\Kalaburagi, Karnataka 585367, India}

\emailAdd{deepaksamuel@cuk.ac.in}
\emailAdd{karthik18495@gmail.com }

\abstract{The India-based Neutrino Observatory (INO) collaboration, as part of its detector R\&D program, has developed prototype stacks of resistive plate chambers (RPCs) to study their performance. These stacks have also been used as testbenches for the development of related hardware and software. A crucial parameter in the characterisation of these detectors and other physics studies is the detection efficiency, which is estimated from track fitting of cosmic muons passing through the stack. So far, a simple straight line fit was used for track fitting, which was sensitive to noise hits and led to rejection of events. In this paper, we present our first results of using  artificial neural networks (ANN) for track fitting of cosmic muons traversing a stack of RPCs. We present in detail, the simulation framework designed for this purpose and show that ANN offers better track reconstruction efficiency than straight line fitting. We also discuss the influence of noise and detection efficiency of cosmic muons on the track reconstruction efficiency.}

\begin{document}
\maketitle
\flushbottom

\section{Introduction}
\label{sec:intro}
The India-based Neutrino Observatory (INO) is a proposed mega-science project to study neutrinos using a 50 kton magnetised iron-calorimeter (ICAL) detector \cite{mondal2012india}. The final detector setup will host approximately 30,000 resistive plate chambers (RPC) as particle trackers. To this end, many detector R\&D programs were initiated and prototype stacks were built to study the RPC parameters using cosmic muons. The detector parameters like efficiency and strip multiplicity are estimated by fitting a straight line to the cosmic muon tracks. Many physics studies performed with this detector also use this simple straight line fitting (SLF) algorithm \cite{pal2012measurement, majumder2012velocity}. This fitting algorithm is straight-forward and robust in many cases but fails when the noise hits in the detector dominate. Though the RPCs will undergo strict quality assurance procedure before deployment and a high noise rate is likely to be an exception than a rule, we explored potential alternatives to the SLF, which are immune to noise hits. In this paper, we present the results of our study of using artificial neural networks (ANN) for fitting cosmic muon tracks through stacked RPCs. We have used the prototype stack at Tata Institute of Fundamental Research (TIFR), Mumbai, as a model for this simulation study. First, we present the details of the detector setup and explain the features of the conventional SLF. We then illustrate in detail the implementation of the ANN and compare the results with those of the SLF by using a simulation framework.

\section{The TIFR Prototype Stack}
\begin{figure}[htbp]
	\centering
		\includegraphics[scale=0.5]{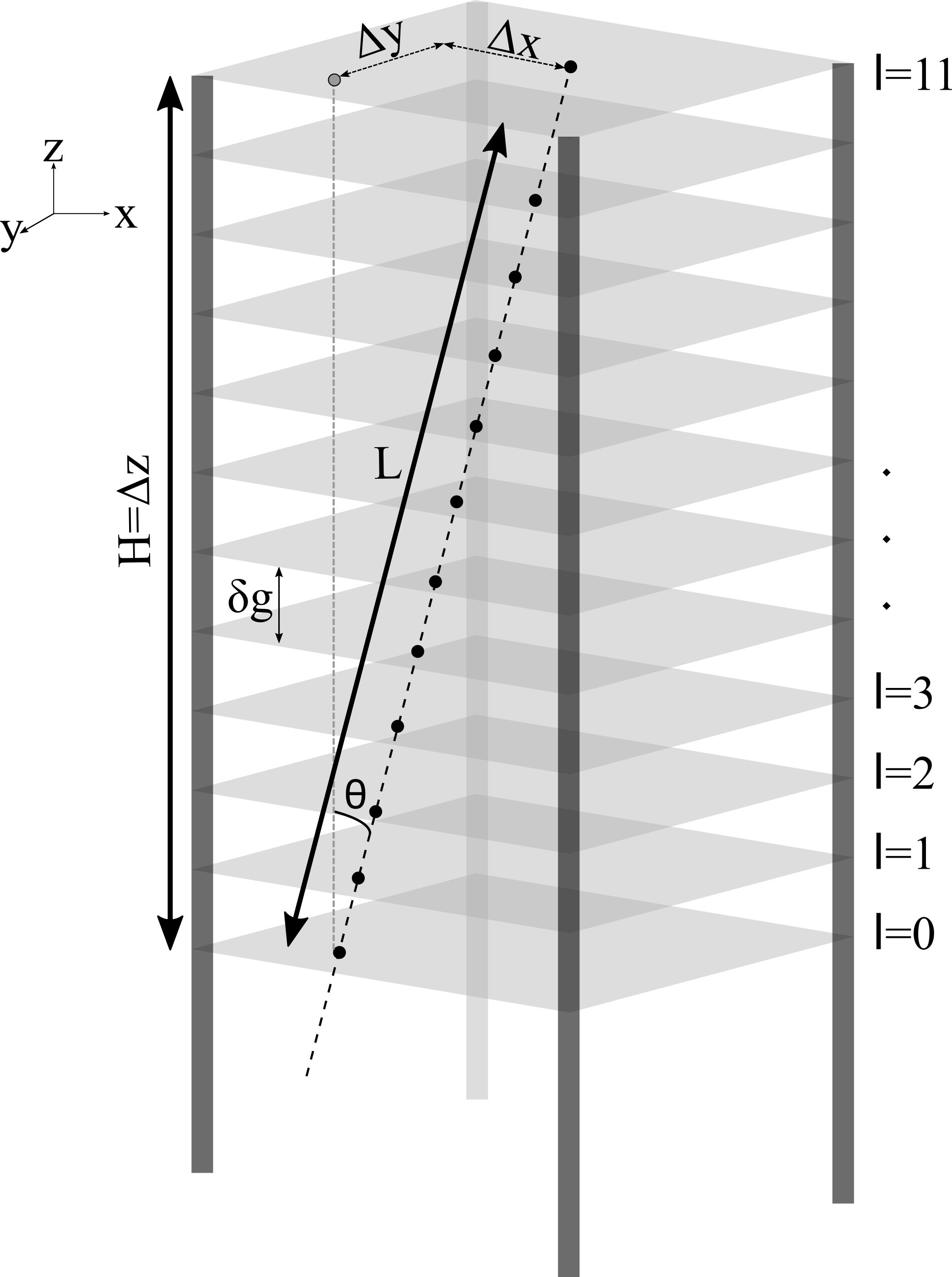}
	\caption{Schematic of the TIFR prototype stack with 12 layers of RPCs.  Cosmic muons traverse along a straight line through the stack. The data acquisition system records the X-Y hits, timing signals and noise rates from the RPCs. Also shown is the zenith angle $\theta$ of the  muon track. Figure adapted from Ref. \cite{angres}.}
	\label{fig:stack}
\end{figure}

RPCs are gas-based detectors consisting of two parallel electrodes of glass or bakelite. The gap between the electrodes is filled with a gas mixture and the sides of the electrodes are sealed. Pick-up panels containing strips of a conducting material like copper are placed over (called X-side) and below the electrodes (called Y-side). The strips are arranged such that the ones on the top are orthogonal to the ones in the bottom, thus enabling the readout of the X-Y hit position of a particle interaction in the gas volume. RPCs are known for their good spatial and temporal resolution and coverage of large areas. A review of RPC construction and features can be found in Refs. \cite{santonico1981development} and  \cite{fonte2002applications}.

The TIFR prototype stack is made of 12 layers of glass RPCs of dimension 1 m $\times$ 1 m stacked on top of each other as shown in figure \ref{fig:stack}. The distance between two RPC layers is 16.8 cm. The glass electrodes of thickness 3 mm are separated by a gas gap of about 2 mm. A conductive coating is applied on top of the glass plates and a high voltage of 10 kV is maintained across the electrodes. The RPCs are operated in the ``avalanche'' mode with a gas mixture of Freon, Isobutane and SF$_6$ in the ratio 95.5 : 4.2 : 0.3. Each pick-up panel has 32 strips each of about 28 mm width interspersed by a gap of about 2 mm.

A VME-based data acquisition system records the X-Y hit positions and timing signals from the RPCs and periodically reads the noise rate from the strips. For most detector or physics studies (and for the present work), the trigger logic is a coincidence of top, middle and bottom layers and hence all tracks must pass through the top and the bottom layer. A detailed description of the detector setup and data acquisition system at TIFR can be found in Refs. \cite{behere2013electronics} and \cite{bhuyan2012vme}.

The design goals of the RPCs with a hit efficiency of > 95 $\%$ and a timing resolution better than 1 ns have already been reached and new developments like testing of 2 m $\times$ 2 m RPCs and fast electronics are being currently pursued.

\section{Track Reconstruction of Cosmic Muons: Straight Line Fit}
\begin{figure}[t!]
	\centering
		\includegraphics[scale=0.25]{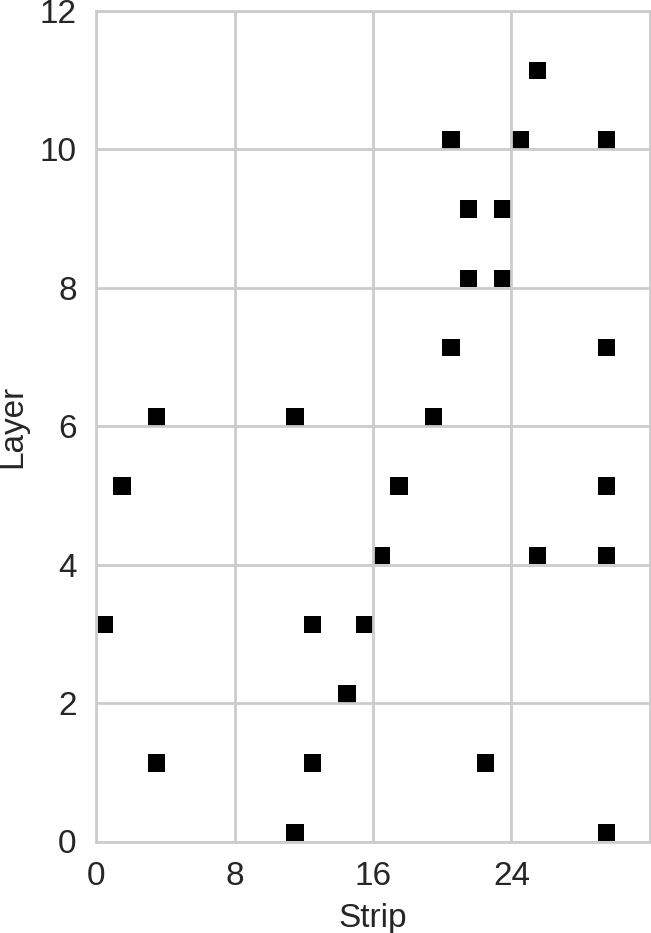}
	\caption{A sample simulated cosmic muon track along with other additional hits. The actual track is 3-dimensional and this view shows only one of the projections. This event will be rejected in the SLF algorithm during the preconditioning stage itself.}
	\label{fig:fail}
\end{figure}

The prototype stacks have been primarily used to study detector parameters like efficiency, noise rate, gas leak rate, response to ambient conditions, changes in gas mixtures and high voltage, etc. Most of the parameters directly or indirectly affect the tracking efficiency of the detector. The tracking efficiency is defined to be the number of cosmic muon tracks which are properly reconstructed divided by the number of triggers generated by the detector. An appropriate statistical condition (for example, reduced $\chi^2 <1)$ is usually set to classify a track as properly reconstructed or not. When a muon passes through a RPC, only the corresponding strips (on the X and Y sides) that they pass through are expected to produce a signal. A SLF is straightforward to implement in such cases. The SLF provides the slope and intercept of the two projections (i.e, X and Y side strip hits) from which the 3-dimensional trajectory of the particle can be reconstructed. However, the presence of outliers in the form of additional hits in the detector, apart from the ones created by the particle, affects the fit parameters. The additional hits may arise due to detector or electronic noise. Hence, preconditioning the data before SLF is often necessary to remove the outliers or to make the SLF less sensitive to outliers. Based on previous experiences, the following cuts are applied before the SLF:   

\begin{itemize}

    \item Condition a: If a layer has more than 2 hits, it is not  considered for the SLF.
    \item Condition b: If a layer has exactly 2 hits and the hits are separated by a distance equivalent to 2 strip-widths, the layer will not be considered for the SLF. Else, the average of the two hits is taken as the hit position for that layer.
    \item Condition c: Once the above layer-level rejection is done, if the number of accepted layers is less than 5, the event is rejected and the track will be classified as "not reconstructed".
    \item Condition d: If the event is accepted, a SLF is done to the hits obtained using the criteria mentioned above and the fit parameters are used to reconstruct the track.
    \item Condition e: It may happen that in spite of the above cuts, there still might be outliers which bias the fit parameters (also reflected in the $\chi^2$ value). In order to reduce such bias, all hit points that are away from the fitted track by more than 2 strip-widths are removed (while retaining the layer in which the hit is less than 2 strip-widths from the first SLF) and a second SLF is made if the number of remaining layers is more than 2. 
    \item Condition f: If the number of remaining layers is less than 2, the event is rejected and the track is classified as "not reconstructed".
\end{itemize}
\begin{figure}[htbp]
	\centering
		\includegraphics[scale=0.65]{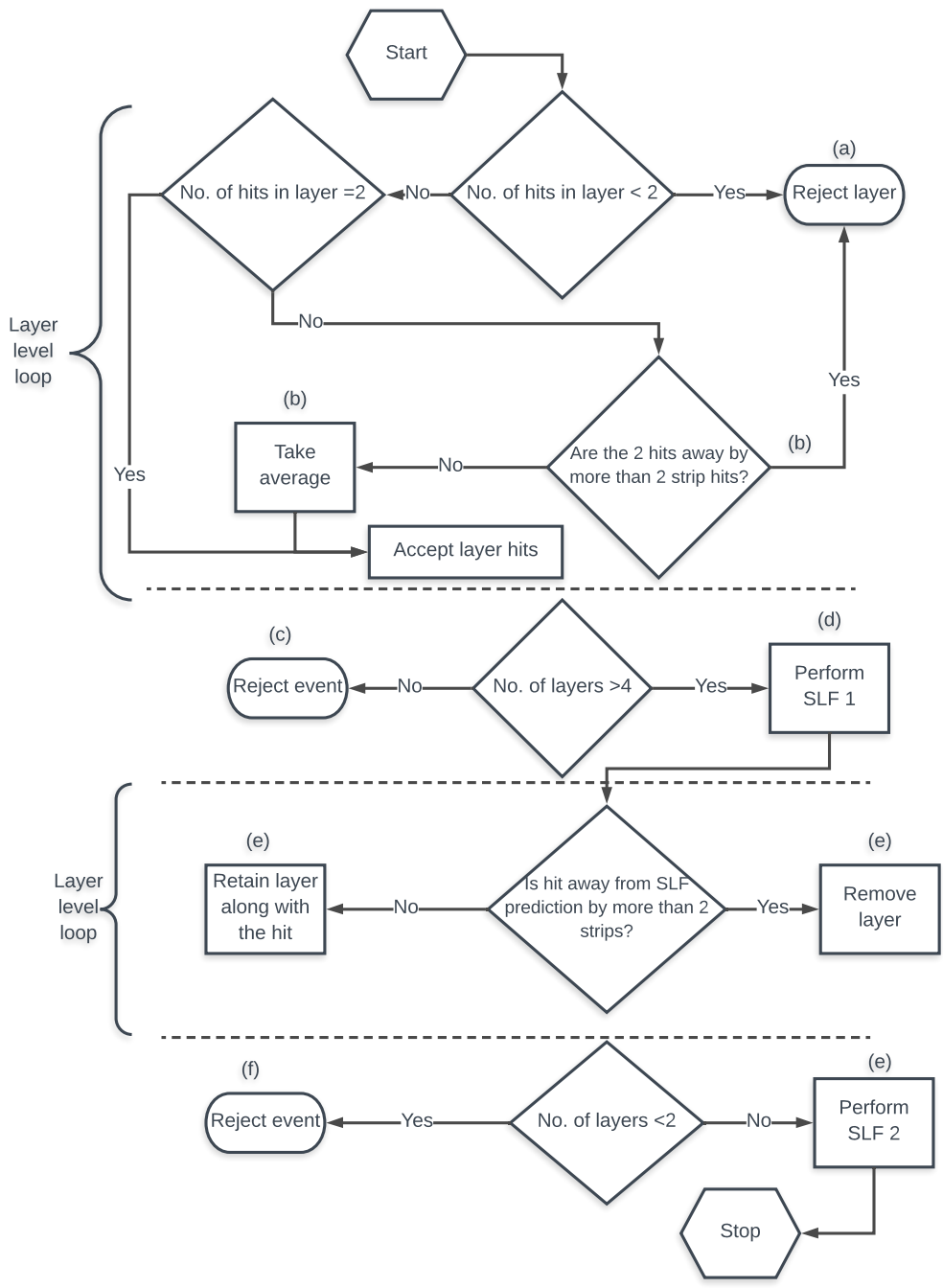}
	\caption{Flowchart showing the sequence of the algorithm used for SLF. The letters indicate the condition that was used to branch.}
	\label{fig:flowchart1}
\end{figure}
The sequence of the algorithm has been explained with a flowchart in figure \ref{fig:flowchart1}. The above cuts have been tested and the performance of the SLF after this filtering has been satisfactory for most of our analysis. However, the SLF fails in cases where more noise hits are present alongside the actual hits, which we shall demonstrate using a simulated event shown in figure \ref{fig:fail}. The figure shows a sample cosmic muon track (simulated) along one of the projections. Table \ref{table1} shows a summary of the accepted and rejected layers after the preconditioning cuts described above. It is evident that since 8 layers are rejected, the event is rejected in accordance with condition c and the track is tagged as "not reconstructed".   Nevertheless, a visual survey of the track shows that there is clearly a track amidst the noise hits. What makes the SLF fail where the visual identification of the track is possible without much effort was the motivation behind this work.

\FloatBarrier

\begin{table}[t!]

\centering
\begin{tabular}{lll}
\toprule
Layer & Accepted/Rejected & Condition \\ \midrule
11& Accepted &  single hit \\ 
10 &Rejected  &  a\\ 
 9&Accepted  &  2 hits away by less than 2\\
 8&Accepted  &  2 hits away by less than 2\\
 7&Rejected  &  b\\
 6&Rejected  &  a\\
 5&Rejected  &  a\\
 4&Rejected  &  a\\
 3&Rejected  &  a\\
 2&Accepted  &  single hit\\
 1&Rejected  &  a\\
 0&Rejected  &  b
\end{tabular}
\caption{Accepted and rejected layers according to the preconditioning for the SLF for the event shown in figure \ref{fig:fail}. }
\label{table1}
\end{table}
\FloatBarrier

\section{Track Reconstruction of Cosmic Muons using ANN}
An overview of the working principle of ANNs is available in Ref. \cite{ann4}. The use of ANNs in the field of high energy physics has already been explored and a wide spectrum of literature is available in this area. More specifically, ANNs have been used for particle identification, track fitting and trigger generation \cite{Wilk:2010pha, ann2, ann3}. In this work, we focus on use of ANNs in cosmic muon track reconstruction in stacked RPCs. \\

\subsection{Simulation framework}
A machine learning algorithm requires two steps - training and testing. In the training step, we feed into the ML program, a dataset which contains the input features and the output labels. In our case, the input features are the strip hit patterns and the outputs are the slope and intercept.
It is useful to visualise the strip hit patterns as a 12 $\times$ 32 matrix with ones at locations where the strip was hit and zeros elsewhere (Ref. figure \ref{fig:fail}). Therefore, for every event, there are two matrices - one for the X-side strip hit patterns and another for the Y-side patterns. The training dataset is now a list of feature matrices each associated with two labels (i.e, the slope and intercept). 

A single pristine event (i.e, without the noise hits) in the dataset is created by randomly generating  slope and intercept values with the constraints imposed by the trigger logic. The conventions in the SLF is such that the slope is in units of strips per layer and the intercept refers to the strip hit in the $0^{th}$ layer. Following the trigger logic used for this work, the slope is constrained between $\pm 2.8$ strips/layer and the intercept is constrained between 0 and 31. The hits for layers 1 to 11 are generated using the slope and intercept, rounded off to the nearest integer. Furthermore, to simulate a realistic event, factors like detector efficiency $\eta$ and strip hit multiplicity $M_s$ have to be accounted for. Finally, the random noise hit multiplicity $M_n$, the effect of which is the focus of this study, is to be added to the event. The three parameters $\eta$, $M_s$ and $M_n$ are inputs to the simulation framework.

\subsection{Folding efficiency and noise factors}
To account for the efficiency, a uniform random number R1 between 0 and 100 is generated for every layer. If R1 is greater than $\eta$ (which is a number between 0 and 100), the 1's in the pristine event for that corresponding layer are replaced by 0's.

When a muon passes through the RPC, there is a likelihood that the adjacent strip is fired in addition to the main strip. This effect is parametrised by the strip hit multiplicity $M_s$. From previous characterisation studies, the average strip hit multiplicity has been found to be around 1.5 and therefore we have restricted $M_s$ to have a minimum value of 1 and a maximum of 2. A uniform random number R2 is generated between 0 and 150 for every layer to account for this parameter. If R2 is between 0 and 50, the strip right to the main strip in the pristine event is set to 1 and if R2 is greater than 100 and less than 150, the strip left to the main strip in the pristine event is set to 1. The hit pattern in a layer is left unaltered if the number is greater than 50 and less than 100.

\begin{figure}[t!]
    \centering
	\begin{subfigure}{0.35\textwidth} 
		\includegraphics[width=\textwidth]{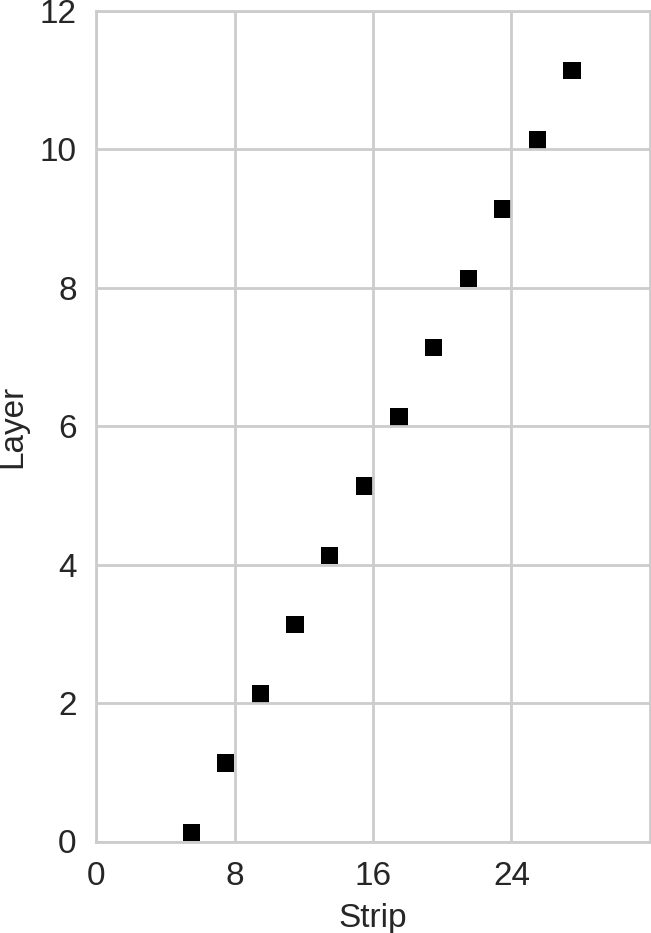}
		\caption{Pristine track without noise hits} 
		\label{fig:pristine track}
	\end{subfigure}
	\hspace{1em} 
	\begin{subfigure}{0.35\textwidth} 
		\includegraphics[width=\textwidth]{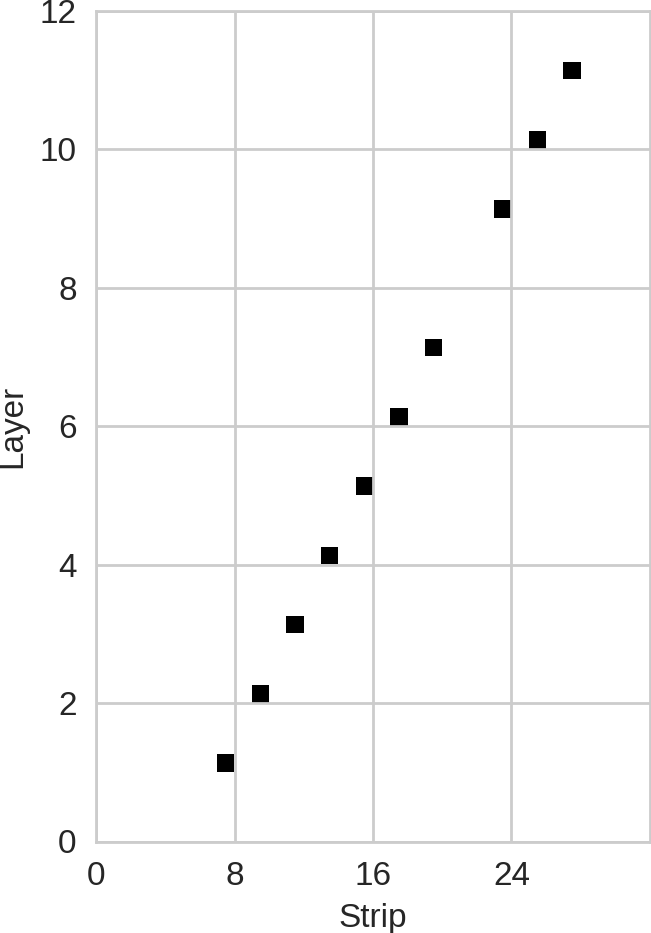}
		\caption{Event (a) modified after including the detector efficiency $\eta$} 
		\label{fig:eff_embed}
	\end{subfigure}
	\hspace{1em}
	\begin{subfigure}{0.35\textwidth} 
		\includegraphics[width=\textwidth]{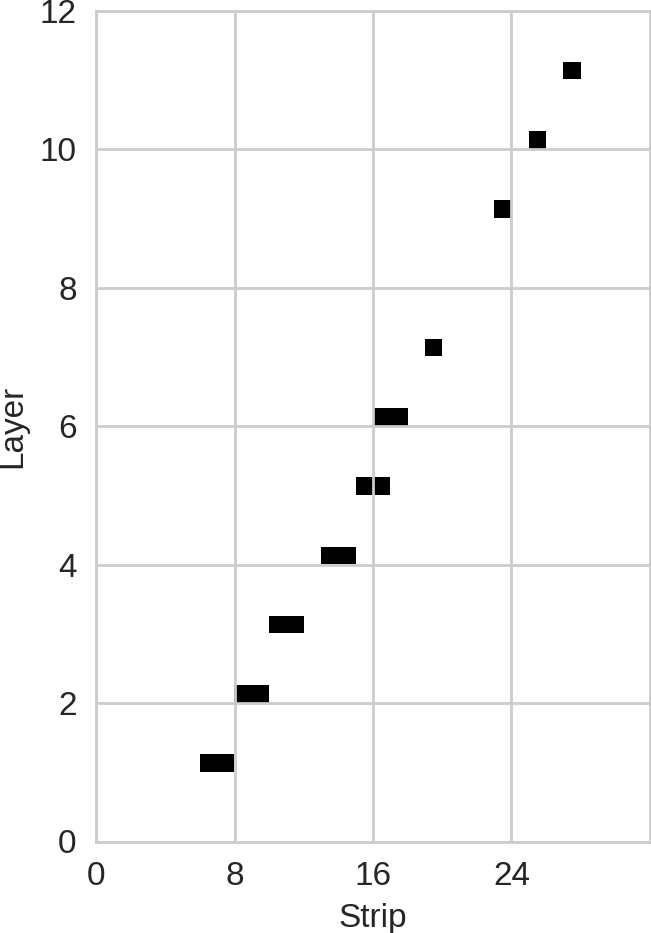}
		\caption{Event (b) after the inclusion of effect of strip multiplicity $M_{s}$} 
		\label{fig:strip_mult}
	\end{subfigure}
	\hspace{1em}
	\begin{subfigure}{0.35\textwidth} 
		\includegraphics[width=\textwidth]{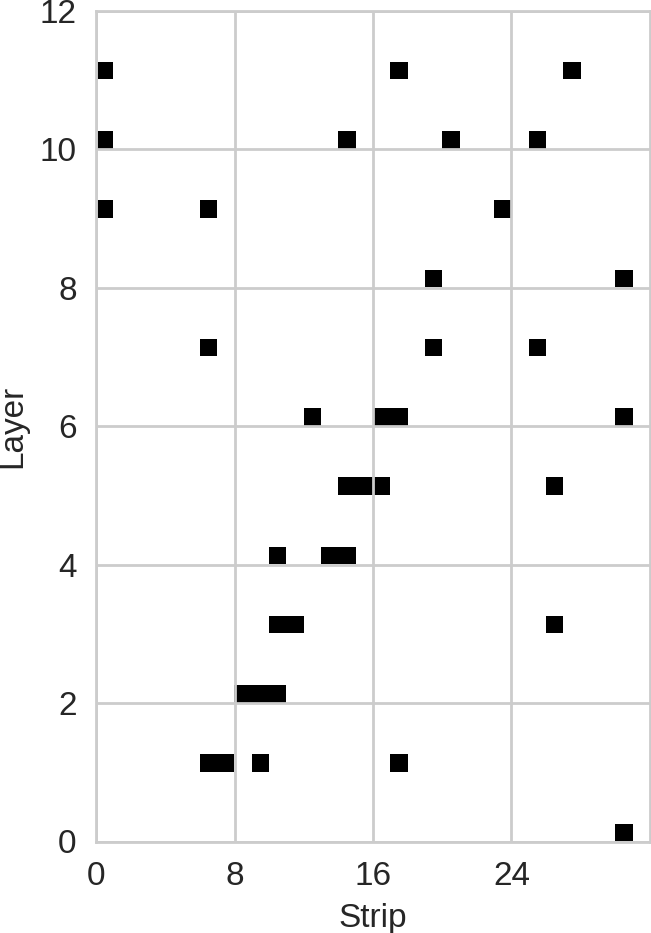}
		\caption{The final track after addition of noise (characterised by the parameter $M_n$) to event (c)} 
		\label{fig:Final_track}
	\end{subfigure}
	\caption{Process of obtaining the final noisy track from a pristine track in the simulation framework.} 
	\label{fig:Simulation_Framework}
\end{figure}

The noise hit multiplicity characterises the maximum number of uncorrelated noise hits in a layer. These may be due to detector noise or electronic noise emanating from different sources like problems in the gas distribution or EMI pickup. To account for this effect, a random number (integer) R3 is generated between 0 and $M_{n}$ for each layer. Then, R3 random integers between 0 and 32 are generated and the corresponding strip numbers are set to 1. 
The sequence of this pattern generation is described with the aid of a sample event shown in figure \ref{fig:Simulation_Framework}.

\subsection{ANN training and testing}
There are a variety of libraries available for machine learning (ML) and we have used scikit-learn, an open-source python based package for our purpose \cite{scikit}. More specifically, the training of the neural network is done using the Multi-layer Perceptron Regressor (MLPRegressor) module in the scikit-learn library. The use of this module is straightforward, however, detailed description of the algorithm is beyond the scope of this paper. More information about the function can be found in Ref. \cite{mlp}. The MLPRegressor function takes as its arguments, the number of hidden layers, the number of neurons in each layer, the activation function to be used, the solver for optimisation, the maximum number of iterations and the tolerance level (the level of loss at which the iterations should stop) for optimisation. The choice of the parameters used in this study was based on trial and error and from the recommendations given in the documentation of the library. We used 4 hidden layers for the slope training each with 120, 320, 12 and 32 neurons, respectively. For the intercept training, we used 2 hidden layers each with 120 and 320 neurons, respectively. The hyperbolic tan function was used for activation and the "adam" solver (recommended for training using large datasets) was used for weight optimisation. The maximum number of iterations was set to 100 and the tolerance to 0.0001. 

The training begins when the function "fit" is called which takes as its input argument, the training features (event matrices) and their associated targets (slope and intercept). The "fit" function returns the optimised biases and weights for the neurons, which are saved in a file. During the testing phase, the "predict" function is called which takes as its argument, a dataset containing the features and returns the associated slope and intercept values.
The generation of dataset for the training and testing of ANN is discussed in the next section.

\section{Data Analysis}
\subsection{Datasets}
\begin{figure}[htpb]
    \centering
    \includegraphics[scale=0.65]{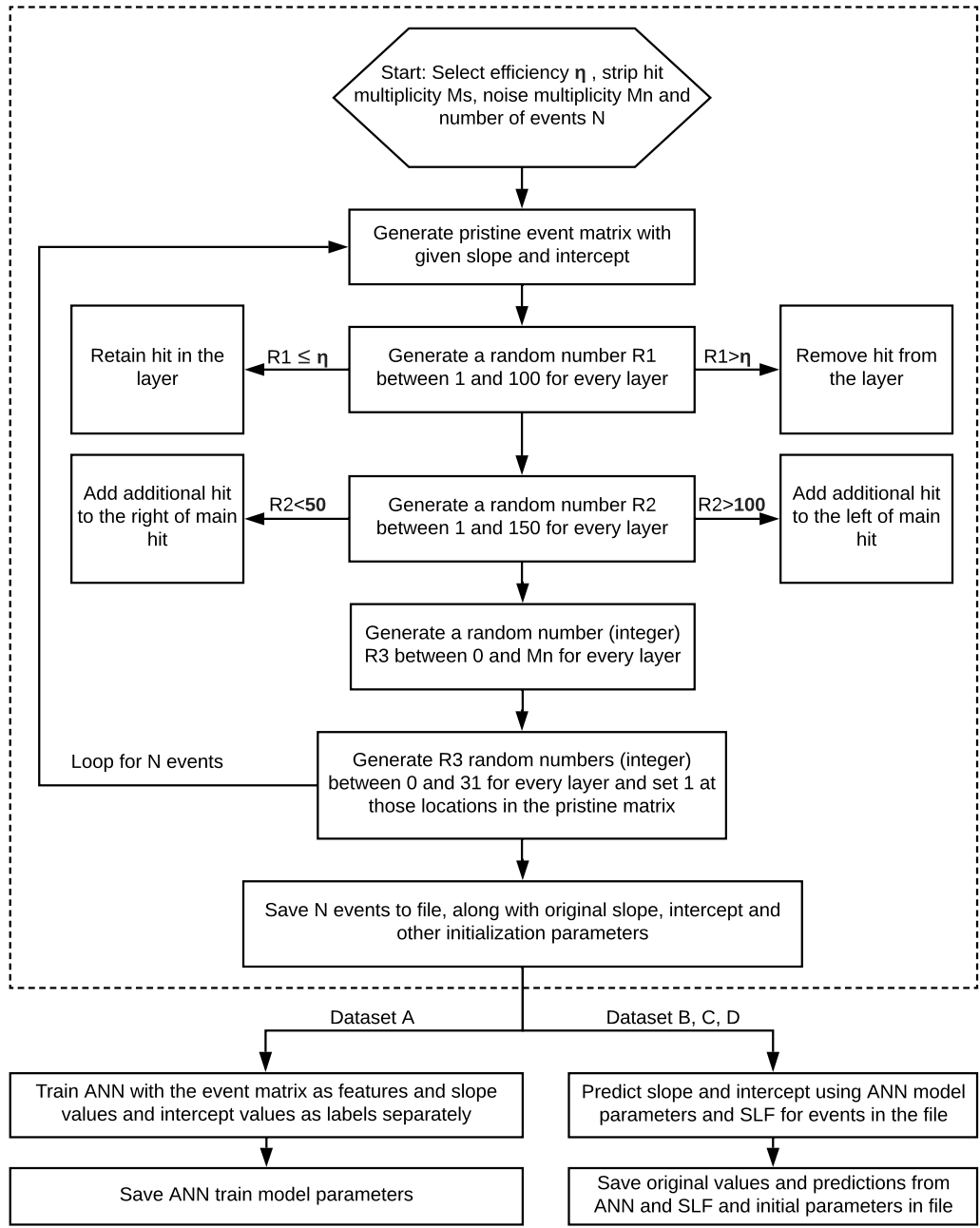}
	\caption{Workflow for dataset generation and the subsequent analysis. The processes inside the dashed box are common for all the datasets.}
	\label{fig:ann_workflow}

\end{figure}

A summary of the datasets generated for the study is shown in table \ref{datasets}. For the training, we used a total of $6 \times 10^6 $  events which were generated with $\eta=100$ and $M_s$=1 and 2. These events were subdivided into 6 classes of $1 \times 10^6 $ events, each class assigned to a specific  noise multiplicity value between 0 and 5 (Dataset A).  To benchmark the performance of ANNs, we generated a pristine dataset with $1 \times 10^5 $ events with $\eta=100\%$, $M_s=1$ and $M_n=0$ (Dataset B). For testing, $6 \times 10^5 $ events with the same input parameters and noise multiplicity  distribution as Dataset A were generated (Dataset C). In addition, for cross-validation, represented by Dataset D, two separate datasets each containing $26 \times 10^6 $ events with $\eta=95\%$ and and an equal number for $\eta=90\%$ were generated. Within the dataset, the number of events was divided such that the multiplicities $M_n$ from 0 to 25 are equally represented. The flow of dataset generation for this study is summarised in figure \ref{fig:ann_workflow}. 
  
\begin{table}[htpb]
\centering
\caption{Datasets generated from the simulation framework. Dataset A and B have 6 subsets of datasets each representing a class of events with an unique noise multiplicity $M_n$ (from 0 to 5). Dataset D has 2 subsets each representing a class of events with an unique efficiency $\eta$. }
\begin{tabular}{llllll}
Dataset & Type &  No. of entries  & $\eta$ & $M_s$ & $M_n$       \\
A & Training & $6\times 10^6$  & 100    & 2      & 0 to 5 \\
B       & Pristine (Bench-marking) & $1\times 10^5$ & 100    & 1      & 0           \\
C       & Testing                 & $6\times 10^5$  & 100    & 2      & 0 to 5 \\
D       & Cross-validation        &  $52\times 10^6$ &  90, 95      &    2    &0 to 25          
\end{tabular}

\label{datasets}
\end{table}

\newpage
\subsection{Benchmarking}

\begin{figure}[htpb]
    \centering
	\begin{subfigure}{0.75\textwidth} 
		\includegraphics[width=\textwidth]{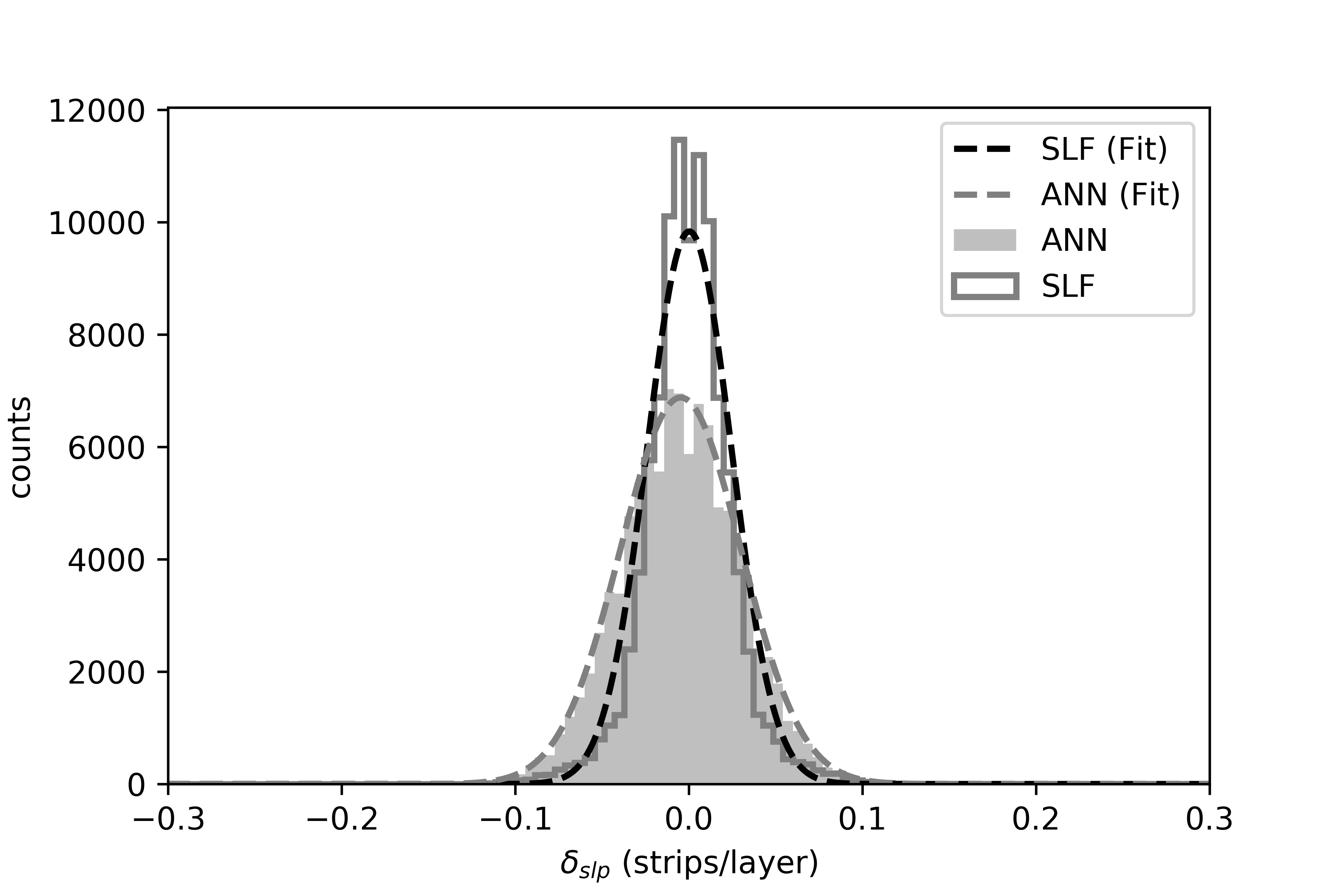}
		\label{fig:residue_slp_pristine}
	\end{subfigure}
	\hspace{0.5em} 
	\begin{subfigure}{0.75\textwidth} 
		\includegraphics[width=\textwidth]{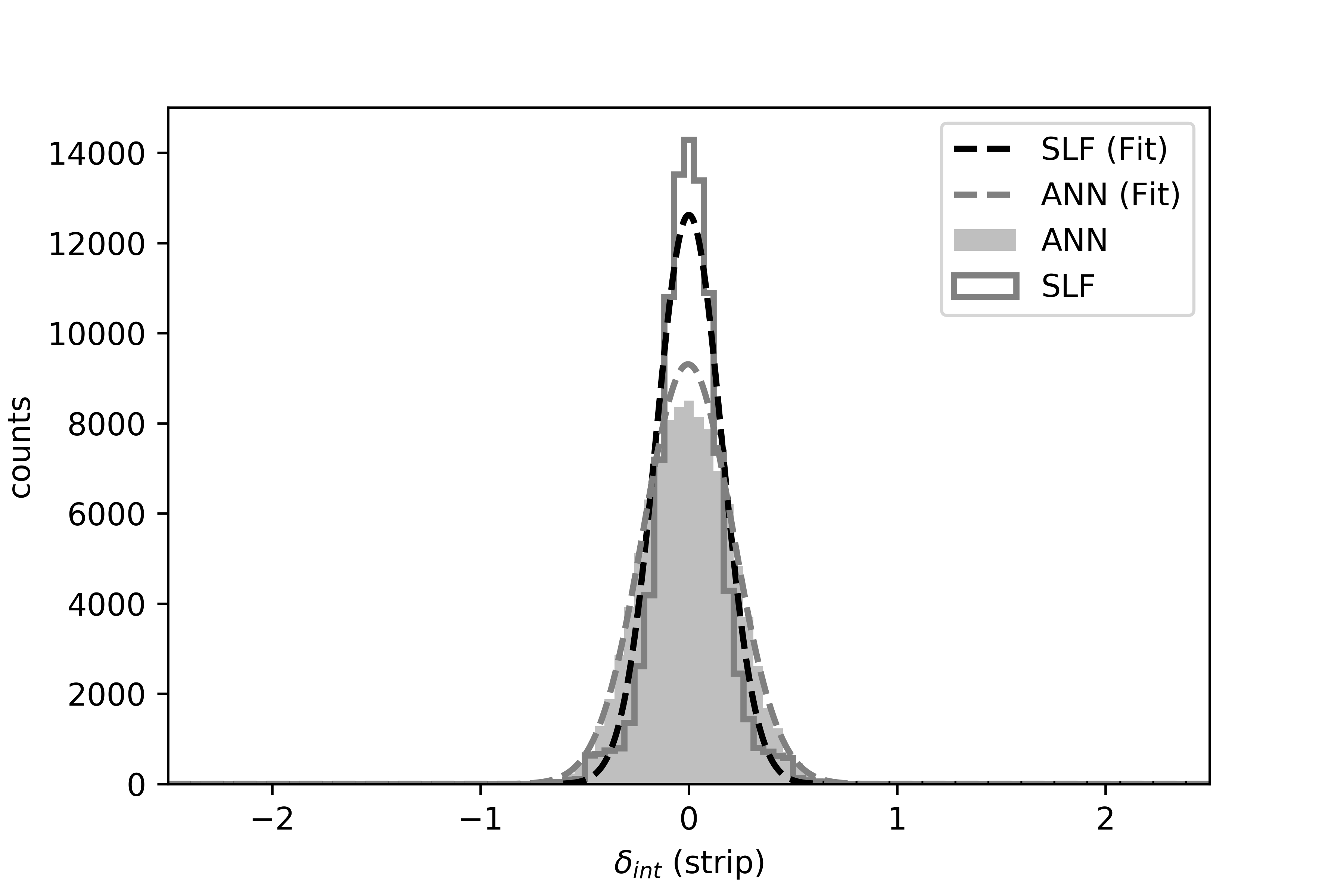}
		\label{fig:residue_int_pristine}
	\end{subfigure}
	\caption{Residual distributions of the slope (top) and intercept (bottom) for the pristine events. The dashed lines represent the gaussian fit to the histograms with the parameters shown in table \ref{fitparams1}} 
	\label{fig:pristine_residue}
	
\end{figure}

  The events in the pristine dataset were simultaneously fit using ANN and SLF to compare their performances with theoretical estimates. The difference between the true slope and the slope predicted by the SLF ($\delta_{slp}^{SLF}$) for these pristine events follows a gaussian distribution, the standard deviation ($\sigma_{slp}^{SLF}$) of which can also be estimated theoretically using error propagation formulas \cite{ku1966notes, bevington2003data}. A detailed derivation of the estimated uncertainties can be found in Ref. \cite{angres}. Figure \ref{fig:pristine_residue} shows this distribution along with the distribution of difference between the true slope and the slope predicted by the ANN ($\delta_{slp}^{ANN}$). The distributions are fitted to a gaussian function, the parameters of which are given in table \ref{fitparams1}.
 
As expected, the standard deviation $\sigma_{slp}^{SLF}$ is close to the expected theoretical value of 0.024 strips/layer. The ANN distribution does not, however, overlap exactly with the SLF distribution. The ANN distribution shows a shorter gaussian (about $30\%$ decrease in the amplitude) with a broader width indicating a slightly higher uncertainty in the estimation of the slope. 
  
A similar distribution can be made for the differences between the true intercept and the intercept predicted by the SLF ($\delta_{int}^{SLF}$) and ANN ($\delta_{int}^{ANN}$). The theoretical estimate for the standard deviation of intercept for pristine events is about 0.14 strips which is close to the standard deviation ($\sigma_{int}^{SLF}$) from the fit parameters. In this case too, the ANN distribution has a larger width compared to the SLF distribution as shown in figure \ref{fig:pristine_residue}, which is also reflected in the fit parameters in table \ref{fitparams1}.
\begin{table}[htpb]
\centering
\caption{Gaussian fit parameters for the pristine slope and intercept distributions shown in figure \ref{fig:pristine_residue}. }

\begin{tabular}{llll}
                & Amp          & Mean                        & Std. Dev                      \\
                & (counts)     & (strips/layer or strips)    & (strips/layer or strips)      \\
Slope (ANN)     & $6879\pm26$  & $(-4.6\pm0.1)\times10^{-3}$ & $(347.9\pm0.8)\times10^{-4}$  \\
Slope (SLF)     & $9837\pm38$  & $(0.3\pm0.1)\times10^{-3}$  & $(243.3\pm0.5)\times10^{-4}$  \\
Intercept (ANN) & $9309\pm36$  & $(-3.5\pm0.7)\times10^{-3}$ & $(2142.7\pm4.8)\times10^{-4}$ \\
Intercept (SLF) & $12622\pm48$ & $(0.0\pm0.5)\times10^{-3}$  & $(1580.3\pm3.5)\times10^{-4}$
\end{tabular}

\label{fitparams1}
\end{table}

\subsection{Residual distributions for higher noise hit multiplicities}
\begin{figure}
    \centering
	\begin{subfigure}{0.75\textwidth} 
		\includegraphics[width=\textwidth]{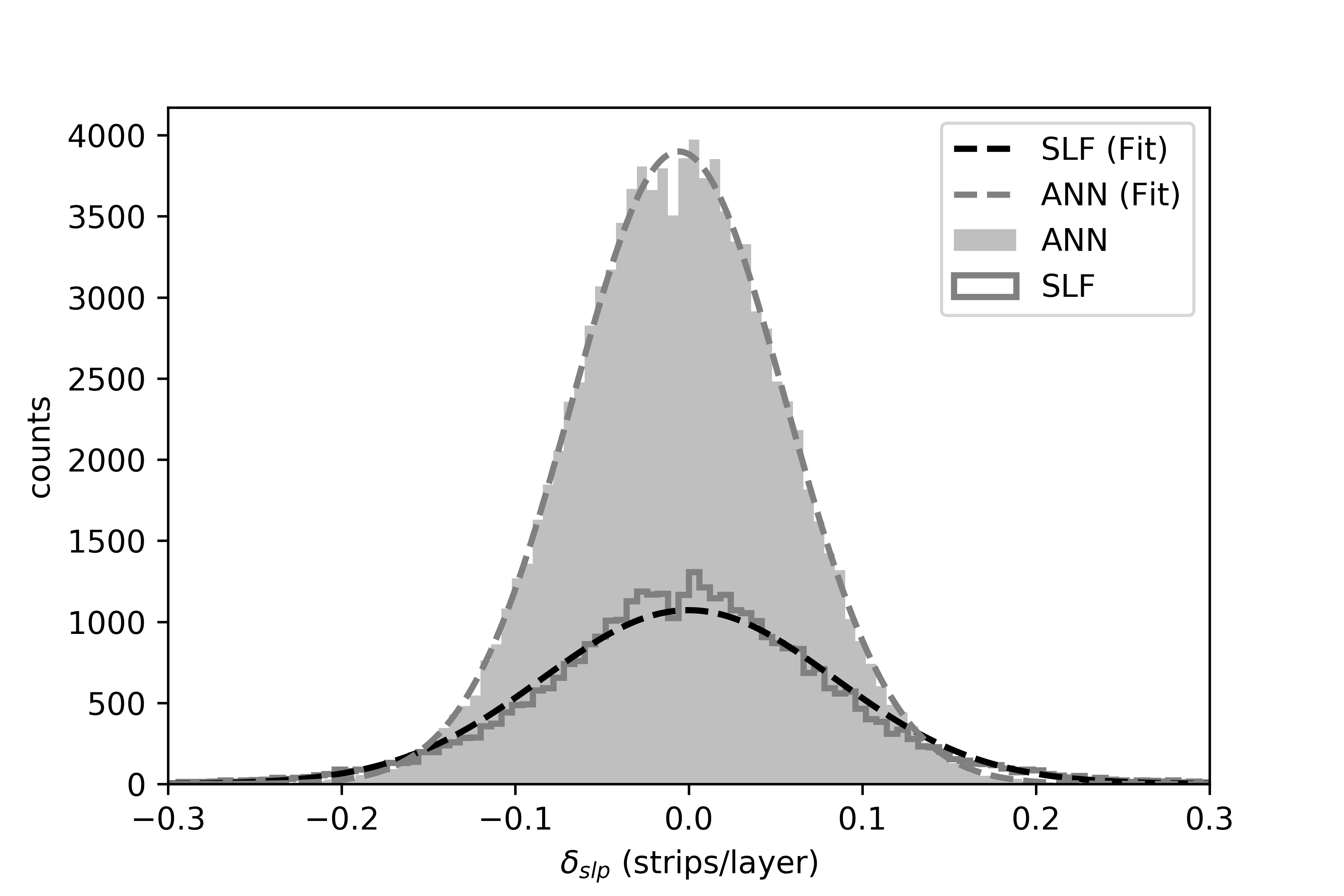}
		\label{fig:residue_slp_mult3}
	\end{subfigure}
	\hspace{0.5em} 
	\begin{subfigure}{0.75\textwidth} 
		\includegraphics[width=\textwidth]{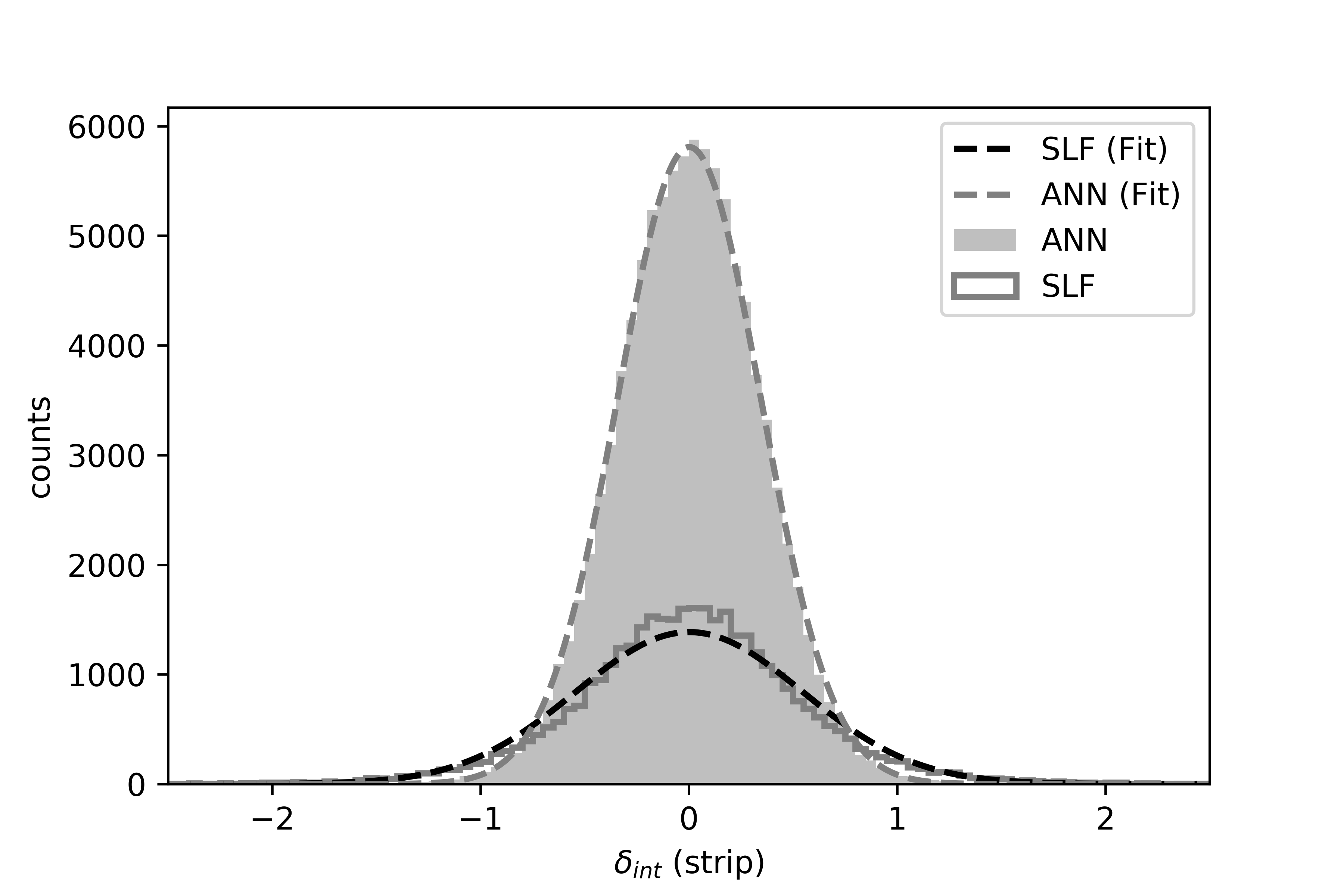}
		\label{fig:residue_int_mult3}
	\end{subfigure}
	\caption{Residual distribution of slope (top) and intercept (bottom) for noisy events with $M_{N}=2$. The dashed lines represent the gaussian fit to the histograms with the parameters shown in table \ref{fitparams2}.} 
	\label{fig:higher_mult_residue}
\end{figure}

The efficacy of ANN in track reconstruction is revealed in events with higher noise hit multiplicities. The events in Dataset A with $M_n$=2 are used to demonstrate this effect and figure \ref{fig:higher_mult_residue} shows the difference distributions (i.e, $\delta_{slp}^{SLF}$,  $\delta_{slp}^{ANN}$, $\delta_{int}^{SLF}$ and $\delta_{slp}^{ANN}$) for this case. The plots show that the height of the SLF distribution is considerably lower than that of the ANN distribution along with an increased spread. This is expected as the probability of rejection of events and the error of the estimated parameters increase in SLF with increasing noise hits. Moreover, the standard deviations of the linear fit parameters have increased by a factor of about 3.5 whereas those of ANN distributions have increased only by a factor of about 2.

\begin{table}[h]
\centering
\caption{Gaussian fit parameters for the slope and intercept distributions for noisy events ($M_n=2$) shown in figure \ref{fig:higher_mult_residue}.}
\begin{tabular}{llll}
                & Amp         & Mean                        & Std. Dev                       \\
                & (counts)    & (strips/layer or strips)    & (strips/layer or strips)       \\
Slope (ANN)     & $3899\pm15$ & $(-5.7\pm0.2)\times10^{-3}$ & $(613.7\pm1.4)\times10^{-4}$   \\
Slope (SLF)     & $1071\pm6$  & $(-0.1\pm0.4)\times10^{-3}$ & $(843.0\pm3.1)\times10^{-4}$   \\
Intercept (ANN) & $5808\pm22$ & $(2.1\pm1.1)\times10^{-3}$  & $(3433.9\pm7.7)\times10^{-4}$  \\
Intercept (SLF) & $1385\pm8$  & $(2.1\pm2.8)\times10^{-3}$  & $(5448.1\pm19.8)\times10^{-4}$
\end{tabular}
\label{fitparams2}
\end{table}

\subsection{Reconstruction efficiency}
The study with the residual distributions in the earlier section elucidates the performance of the ANN algorithm. However, for faithful track reconstruction, both the slope and the intercept have to be correctly estimated. Therefore, to understand the performance of ANN in  reconstructing a track, we study the variation of reconstruction efficiency $\eta_{rec}$ as a function of $M_n$ for different $\eta$. The parameter $\eta_{rec}$ is defined as the ratio of the number of events which have both  $\delta_{slp}$ and $\delta_{int}$ within $\pm3\sigma_{slp}^{SLF}$ and $\pm3\sigma_{int}^{SLF}$, respectively, obtained from theoretical estimates of pristine tracks (i.e, $\pm0.072$ strips/layer and $\pm0.480$ strips) to the total number of events. 

\begin{equation}
    \eta_{rec} = \frac{\text{No. of events with both } |\delta_{slp}| < \pm0.072 \text{ and } |\delta_{int}| < \pm0.480}{\text{Total number of events}}
\end{equation}

\begin{figure}
    \centering
    \includegraphics[scale=0.75]{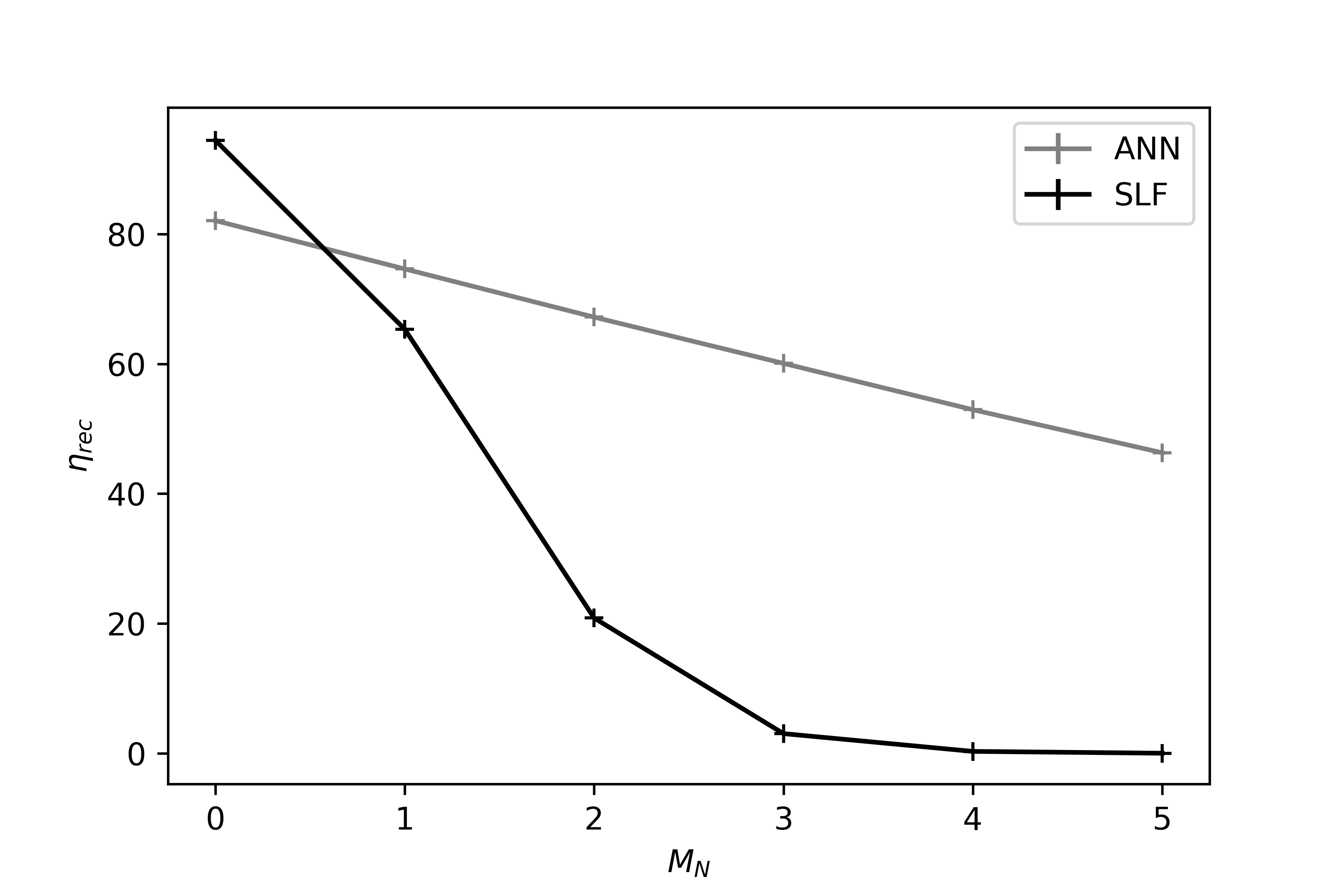}
	\caption{Variation of reconstruction efficiency for the SLF and ANN with noise hit multiplicity $M_n$, generated using the testing dataset (Dataset C). The error due to finite statistics is less than a percent and is not visible in this plot.}
	\label{fig:rec_eff_100}

\end{figure}

Figure \ref{fig:rec_eff_100} shows the variation of $\eta_{rec}$ as a function of $M_{n}$ for an efficiency $\eta$ of 100$\%$. This analysis was done using Dataset C (Testing). The figure shows that, in case of SLF, even with the best detection efficiency, the reconstruction  efficiency drastically falls down close to 20$\%$ for $M_{n}$ = 2. On the other hand, the ANN has a reconstruction efficiency of close to 60$\%$ even with $M_{n}$ = 5 whereas the SLF completely fails to reconstruct in this case. The failure of SLF is, once again, due to the conditions imposed and described earlier, which lead to rejection of events in the preconditioning stage itself. Therefore, one might consider performing an SLF without conditions a and b but the question as to how to estimate the actual hit for layers with multiple hits needs to be addressed. One possibility is to take the average of the hits in a layer and then perform the SLF. Intuitively, one may guess that while this may work for hits which are close to each other, it may also lead to larger errors on the fit parameters when the hits are scattered. 

\begin{figure}
    \centering
	\begin{subfigure}{0.75\textwidth} 
		\includegraphics[width=\textwidth]{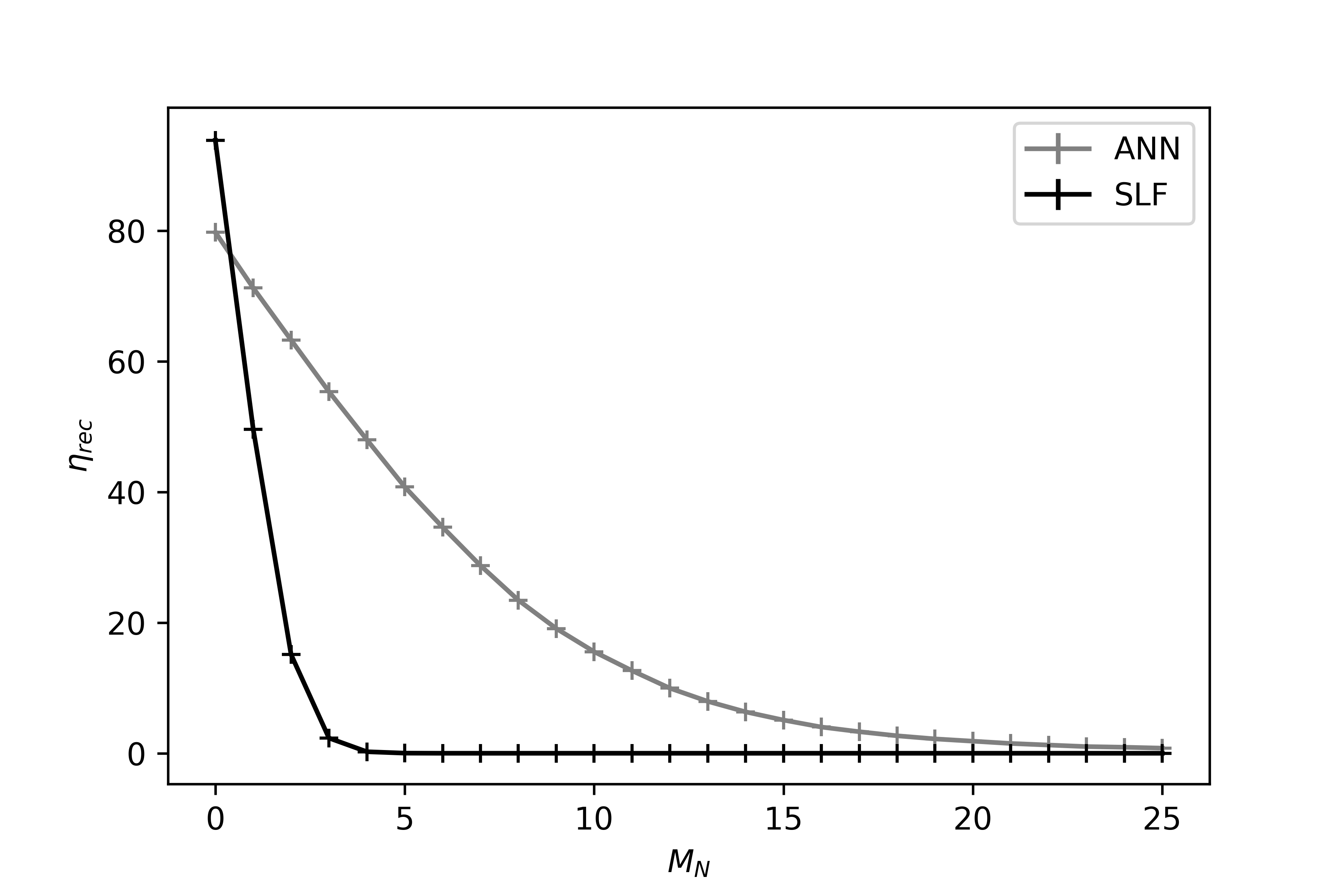}
	\end{subfigure}
	\hspace{0.5em} 
	\begin{subfigure}{0.75\textwidth} 
		\includegraphics[width=\textwidth]{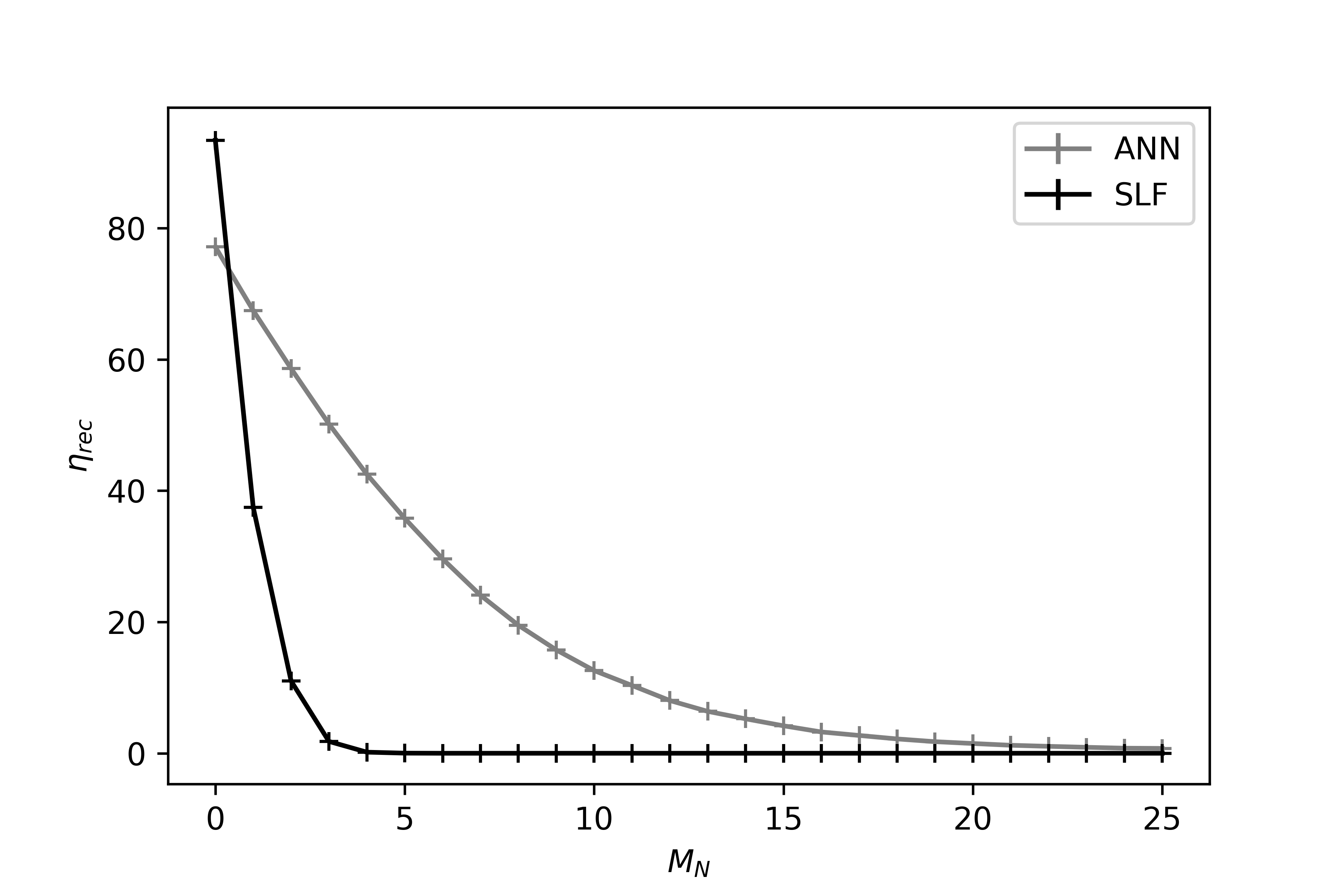}
		\label{fig:rec_eff_90}
	\end{subfigure}
	\caption{Variation of reconstruction efficiency ($\eta_{rec}$) as a function of $M_{n}$ for detector efficiency $\eta=95\%$ (top) and $\eta=90\%$ (bottom). The error due to finite statistics is less than a percent and is not visible in this plot.} 
	\label{fig:rec_eff}
\end{figure}

The predictive performance of ANN can be tested using a similar study but with the cross-validation datasets with $\eta=90$ and $\eta=95$. It is noteworthy to mention here that these events are not similar to the events in training datasets and therefore this is a true test of the functionality of the ANN algorithm. Figure \ref{fig:rec_eff} shows the plot for the two cross-validation datasets that shows a similar trend as the testing dataset. Thus, the ANN is superior to SLF in terms of its predictions in cases with high noise multiplicities.  

There is a common trend that is seen in the reconstruction efficiency generated using the testing and cross-validation datasets - the reconstruction efficiency for ANN is offset to about 80$\%$ at $M_N=0$ whereas the SLF shows a 100$\%$ efficiency. This is due to the fact that in the case of SLF, the events are fit separately and the fit parameters are optimised on an event-by-event basis. On the other hand, in the case of the ANN, the fit parameters are predicted from a model which was trained using a dataset that contained events with different multiplicities ranging from 0 to 5. Therefore, the weights and biases of the model (which are fixed values after training with all these events) are optimised taking all the events into account and in such a way that the overall deviations of predictions from the actual value is minimised. Thus, in minimising the overall loss function to optimise the weights, we compromise on the predictive accuracy for certain class of events. However, this issue can be overcome, if we train and generate a model separately for different multiplicities and use it for prediction for that specific case of multiplicity. This aspect will be explored in our future work.

\section{Computational Resources}
For the work carried out in this study, the code was developed and the datasets were generated on a laptop with Windows 10 operating system and an Intel-i7-7550 CPU running at 2.70 GHz. The laptop was equipped with 8.00 GB physical memory and a 1 TB hard-drive. A brief summary of the time taken for execution of various processes is given in table \ref{timeforproc}. As seen from the table, ANN fitting is faster than SLF as the model parameters are already fixed in the training process of ANN while in the case of SLF, a minimisation process is carried out for every event, increasing the computation time. This could be important for studies on offline triggering systems where event filtering has to be done at a fast rate.   
\begin{table}[htbp]
\centering
\caption{Time taken for generation of different datasets used in this study. For reference, the time taken for retrieving the fit parameters from ANN and SLF for pristine events is given in the last two rows.}
\begin{tabular}{lll}
Process     & Number of events & Approximate time taken (min) \\
Dataset A   & $6\times10^6$    & 30                           \\
Dataset B   & $10^5$           & 3                            \\
Dataset C   & $6\times10^5$    & 15                           \\
Dataset D   & $52\times10^6$   & 120                          \\
ANN fitting & $10^5$           & 0.5                          \\
SLF fitting & $10^5$           & 2                           
\end{tabular}
\label{timeforproc}
\end{table}
\section{Results and Conclusions}
\begin{figure}[t!]
	\begin{subfigure}{0.25\textwidth} 
		\includegraphics[width=\textwidth]{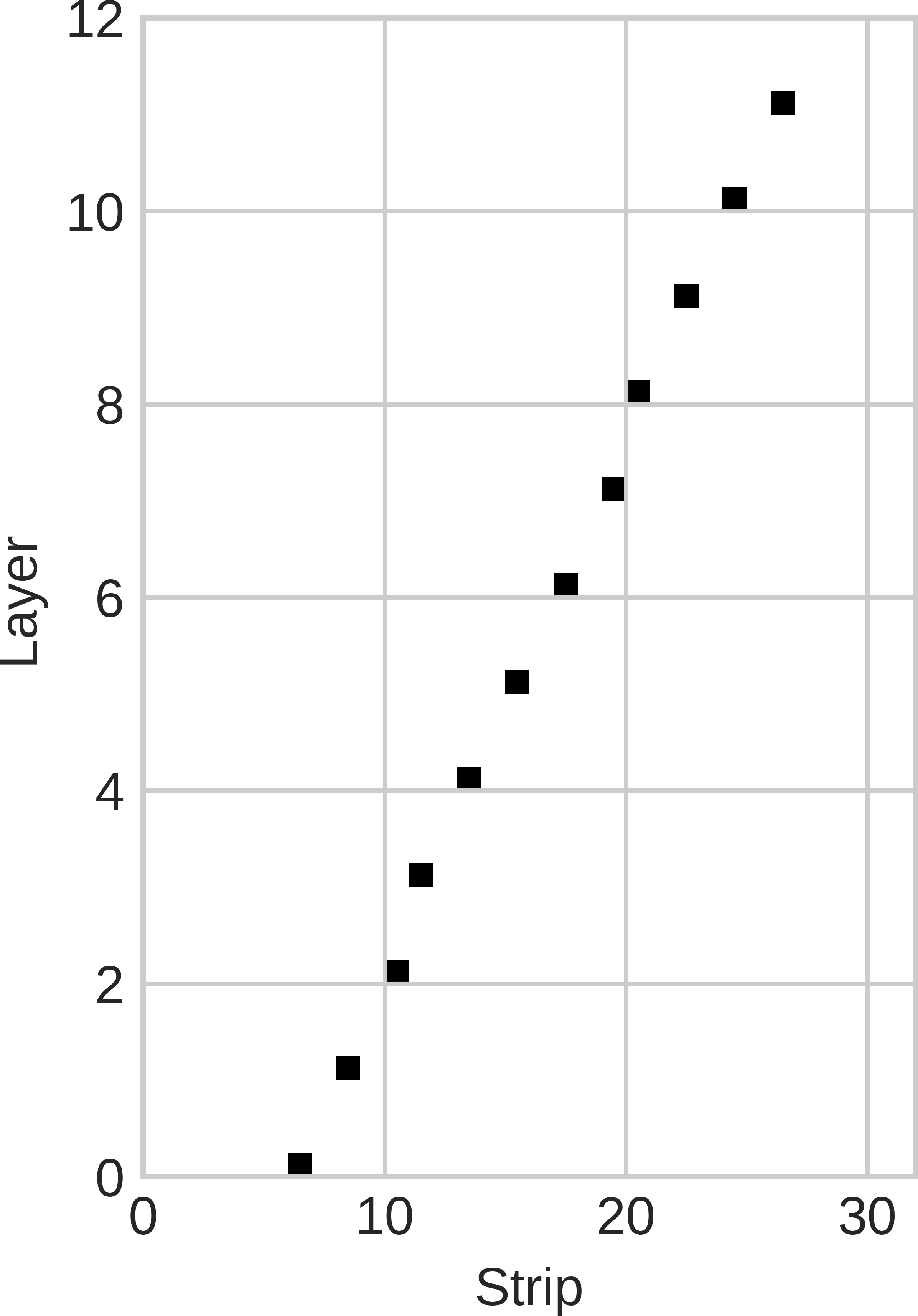}
		\label{fig:rec_1a}
	\end{subfigure}
	\hspace{1em} 
	\begin{subfigure}{0.25\textwidth} 
		\includegraphics[width=\textwidth]{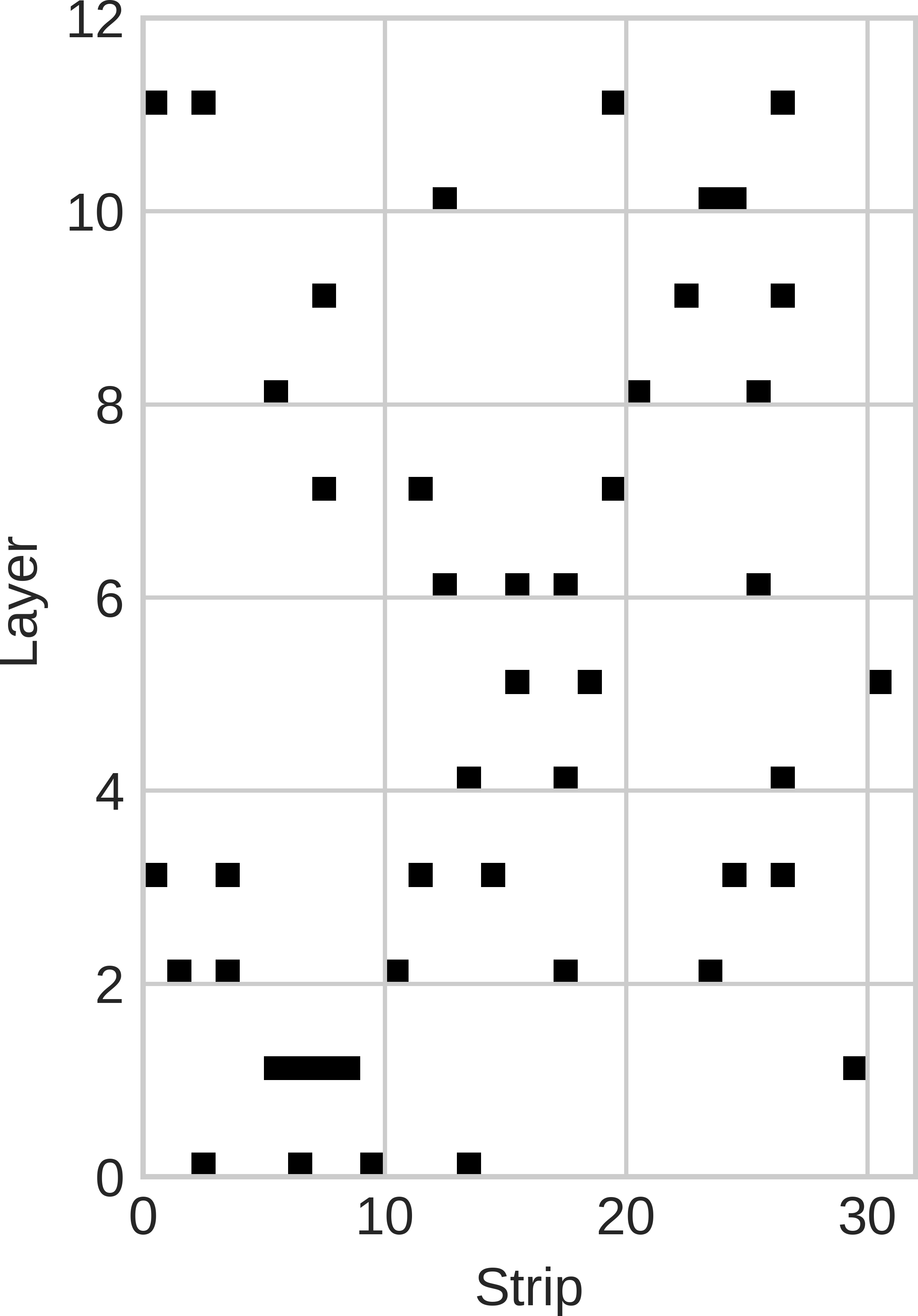}
		\label{fig:rec_1b}
	\end{subfigure}
	\hspace{1em}
	\begin{subfigure}{0.25\textwidth} 
		\includegraphics[width=\textwidth]{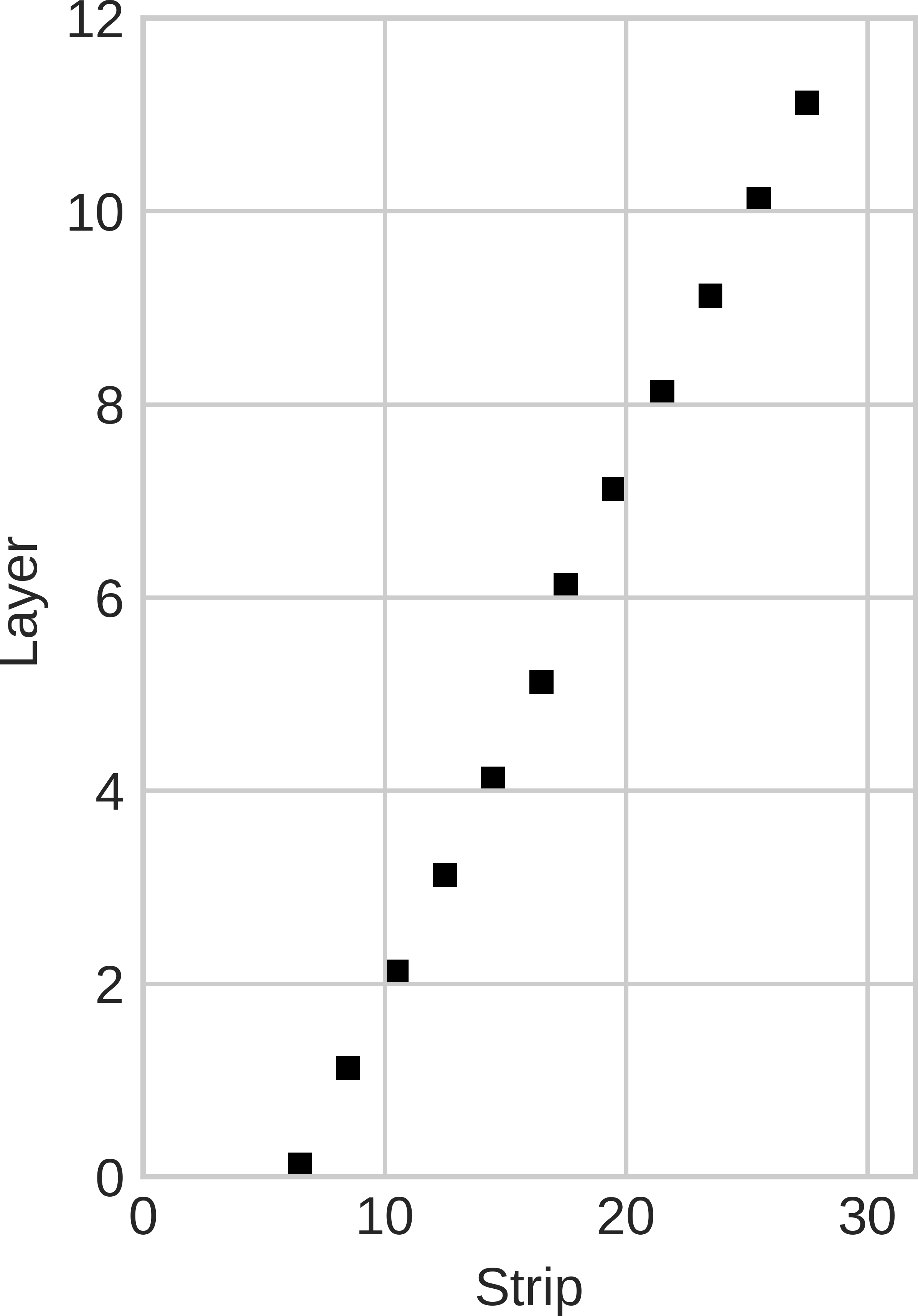}
		\label{fig:rec_1c}
	\end{subfigure}
	\hspace{1em}
	\begin{subfigure}{0.25\textwidth} 
		\includegraphics[width=\textwidth]{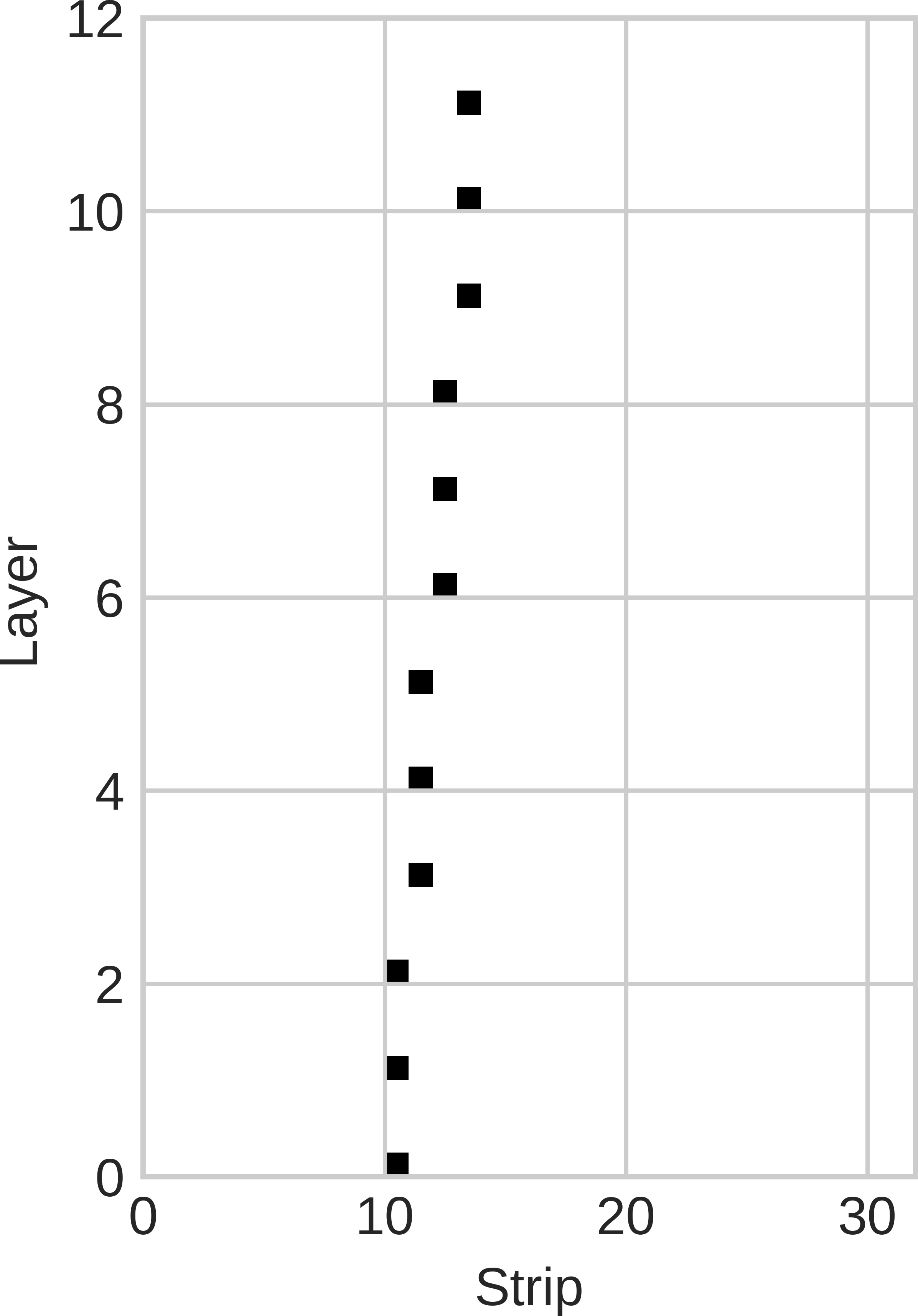}
		\label{fig:rec_2a}
	\end{subfigure}
	\hspace{1em}
	\begin{subfigure}{0.25\textwidth} 
		\includegraphics[width=\textwidth]{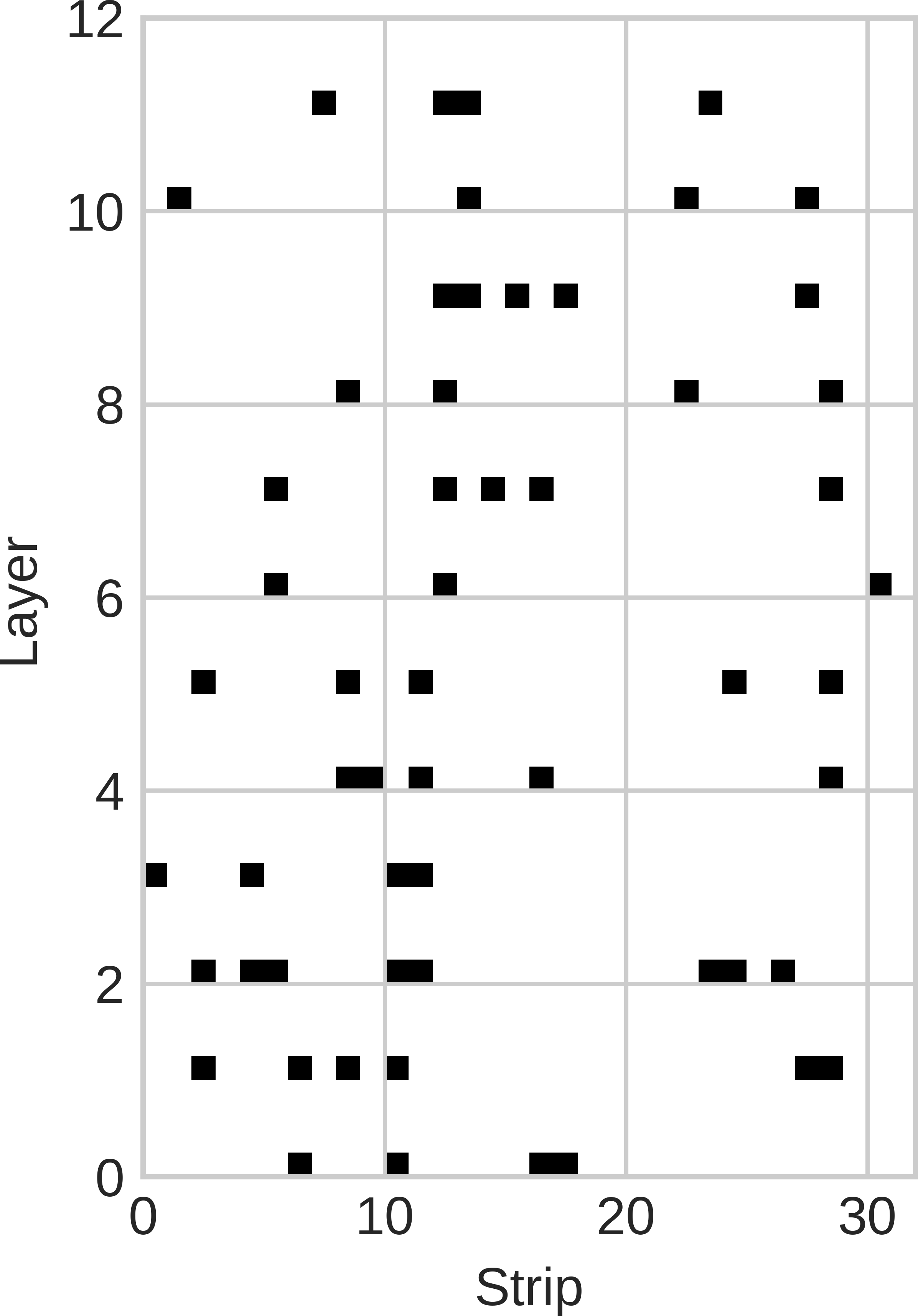}
		\label{fig:rec_2b}
	\end{subfigure}
	\hspace{1em} 
	\begin{subfigure}{0.25\textwidth} 
		\includegraphics[width=\textwidth]{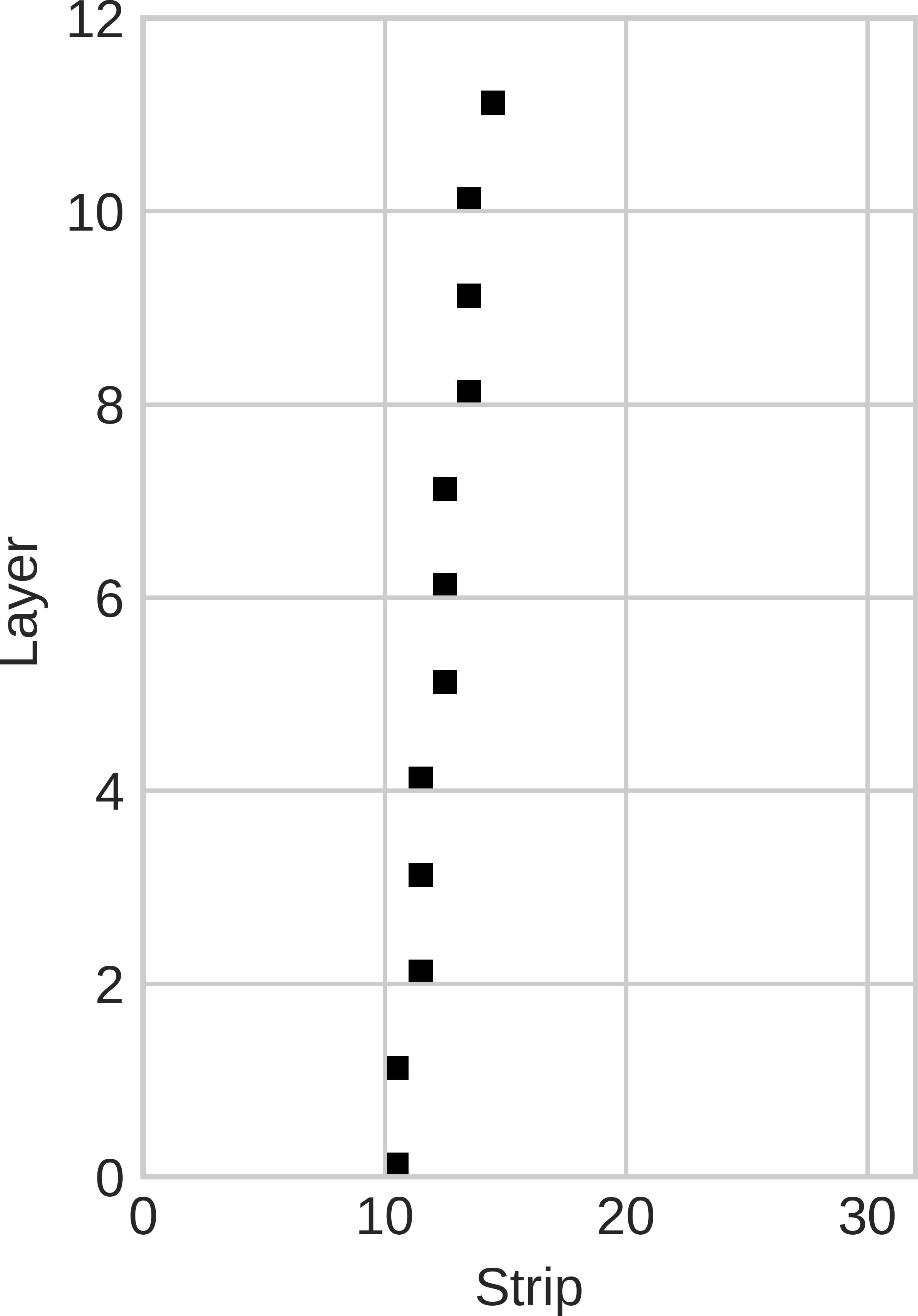}
		\label{fig:rec_2c}
	\end{subfigure}
	\hspace{1em}
	\begin{subfigure}{0.25\textwidth} 
		\includegraphics[width=\textwidth]{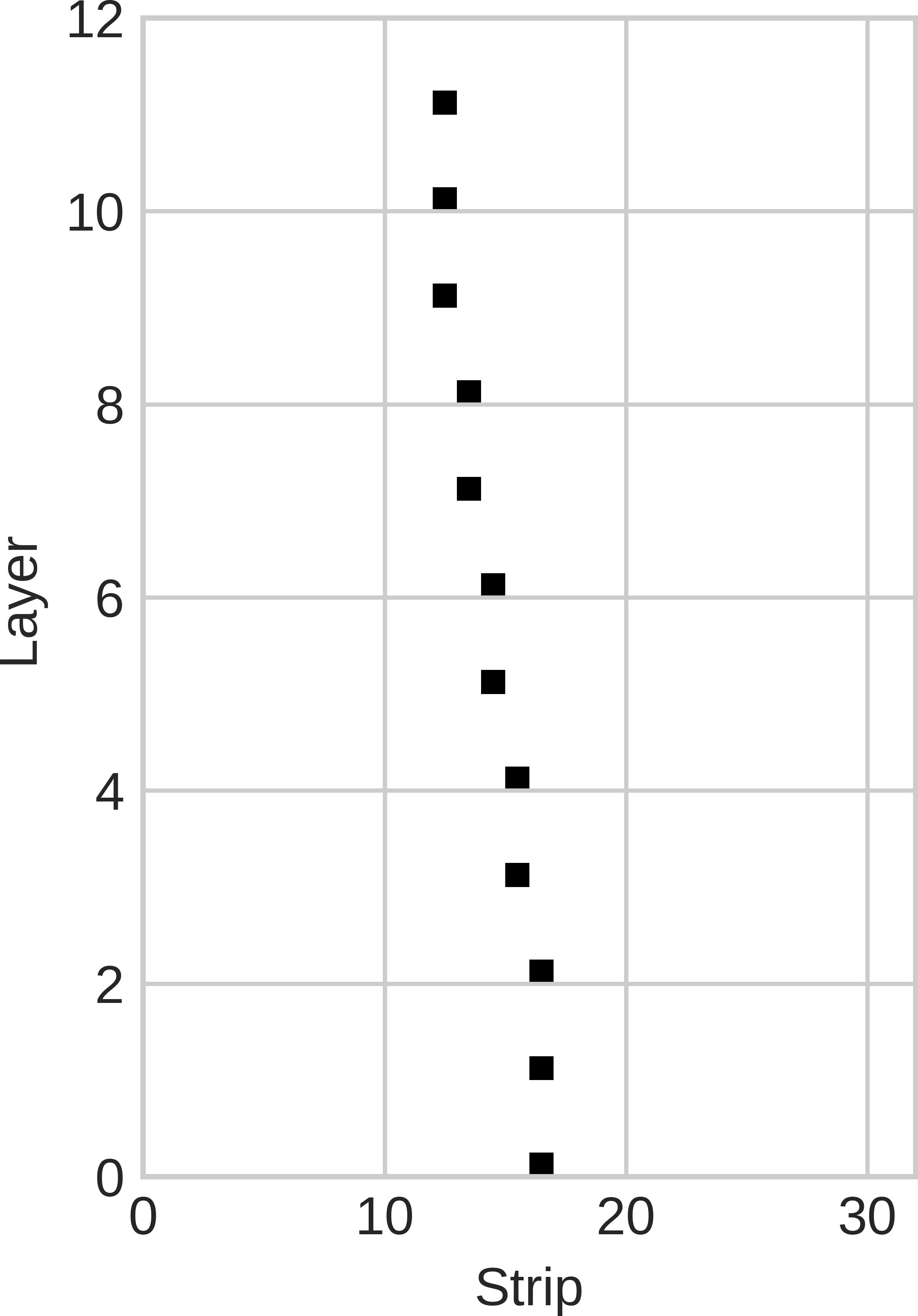}
		\label{fig:rec_3a}
	\end{subfigure}
	\hspace{4em}	
	\begin{subfigure}{0.25\textwidth} 
		\includegraphics[width=\textwidth]{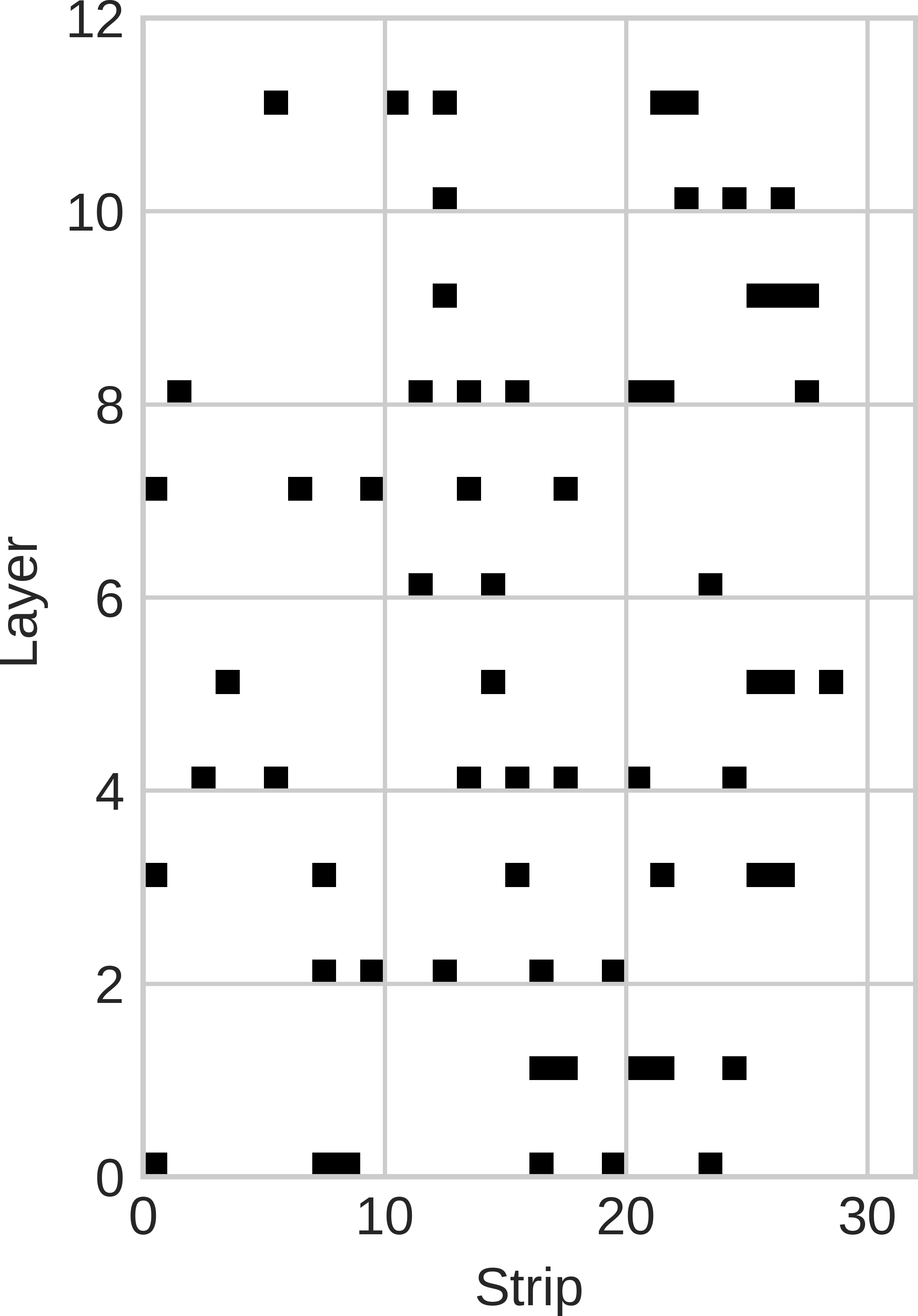}
		\label{fig:rec_3b}
	\end{subfigure}
	\hspace{4em}
	\begin{subfigure}{0.25\textwidth} 
		\includegraphics[width=\textwidth]{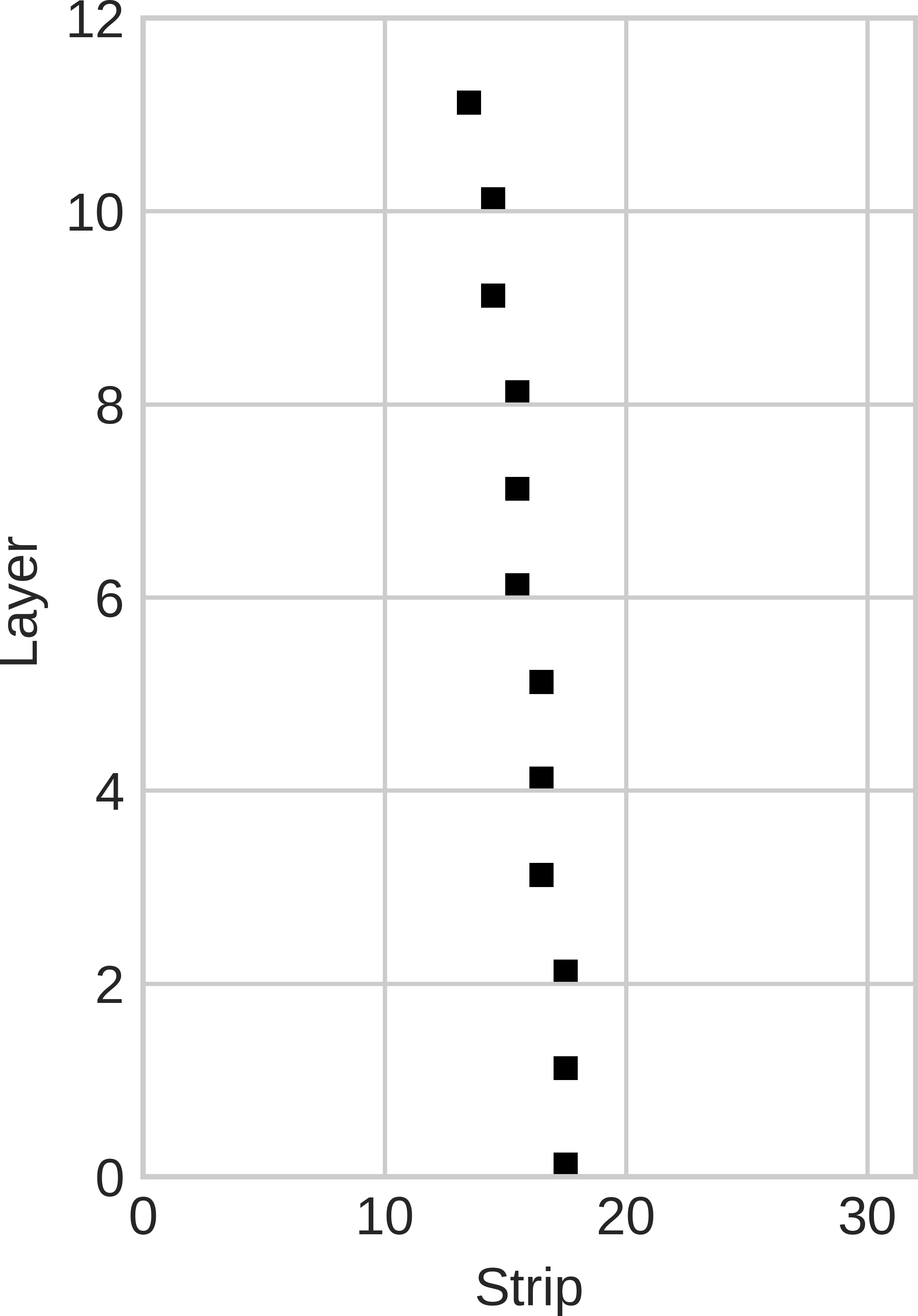}
		\label{fig:rec_3c}
	\end{subfigure}
	\hspace{1em}
	\caption{Few sample reconstructions showing the ability of ANN to track cosmic muons in a noisy environment. The events on the left column are the pristine events, the events in the middle are the pristine events after the inclusion of efficiency and noise hit multiplicity factors. The right column shows the track as predicted by ANN.} 
	\label{fig:Sample Reconstructions}

\end{figure}

In this study, we have presented ANN as a potential alternative to the conventional straight line fitting. We have shown that the predictive performance of ANN, as evidenced by the reconstruction efficiency, is better than that of SLF. We present few exemplary reconstructions in figure \ref{fig:Sample Reconstructions}, which further demonstrate this superiority. As seen from these events, it is clear that ANN outperforms SLF in case of noisy events, which are rejected by the SLF. 

The performance of RPCs often degrade with time due to ageing effects,  which impact the noise rate and efficiency in addition to other parameters \cite{aging}.  There are various methods proposed to delay the ageing of the RPCs and to recover them after ageing \cite{aging2}. However, it might not be always technically feasible to carry out these processes or to replace the detectors immediately. The outcome of this study offers scope to extend the use of detectors which show increased noise rates and decreased detection efficiency. We believe that this study will also provide a foundation for future activities in the INO-ICAL experiment where ANN could be potentially employed, for instance, in track fitting of muons in a magnetic field.

\FloatBarrier

\acknowledgments
The INO project is funded by the Department of Atomic Energy and the Department of Science and Technology, Government of India. The authors would like to thank the scientific and technical staff at IICHEP, Madurai and TIFR, Mumbai for their useful suggestions and comments. DS likes to thank Dr. Sudeshna Dasgupta for her suggestions on improving the manuscript. 



\begin{thebibliography}{99}
\bibitem{mondal2012india}
N. K. Mondal, \emph{India-based neutrino observatory (INO),} \emph{Eur. Phys. J. Plus} {\bf 127.9} (2012) 1.

\bibitem{pal2012measurement}
 S. Pal, et al., \emph{Measurement of integrated flux of cosmic ray muons at sea level using the INO-ICAL prototype detector,} \emph{J. Cosmol. Astropart. Phys.} {\bf  2012.07} (2012) 033.

\bibitem{majumder2012velocity}
G. Majumder, et al., \emph{Velocity measurement of cosmic muons using the India-based neutrino observatory prototype detector,} \emph{Nucl.  Instrum. Methods Phys. Res., Sect. A} {\bf 661} (2012) S77.


\bibitem{santonico1981development}
R. Santonico and R. Cardarelli. \emph{Development of resistive plate counters,} \emph{Nucl.  Instrum. Methods Phys. Res., Sect. A} {\bf 187.2-3} (1981) 377.

\bibitem{fonte2002applications}
P. Fonte, \emph{Applications and new developments in resistive plate chambers,} \emph{IEEE Trans. Nucl. Sci.} {\bf 49.3} (2002) 881.

\bibitem{behere2013electronics}
A. Behere, et al., \emph{Electronics and data acquisition system for the ICAL prototype detector of India-based neutrino observatory,}  \emph{Nucl.  Instrum. Methods Phys. Res., Sect. A} {\bf 701} (2013) 153.

\bibitem{bhuyan2012vme}
 M. Bhuyan, et al., \emph{VME-based data acquisition system for the India-based neutrino observatory prototype detector,}  \emph{Nucl.  Instrum. Methods Phys. Res., Sect. A} {\bf 661} (2012) S73.


\bibitem{ann1}
C. Bortolotto, et al., \emph{Neural networks in experimental high-energy physics,} \emph{Int. J. Mod. Phys. C} {\bf 3.04} (1992) 733.


\bibitem{ann4}  
 R. Lippmann, \emph{An introduction to computing with neural nets,} \emph{IEEE Assp Mag.}  {\bf 4.2} (1987) 4.
  
 

\bibitem{Wilk:2010pha} 
  A.~Wilk,
  \emph{Particle Identification using Artificial Neural Networks with the ALICE TRD,}   \emph{PhD thesis} {\bf CERN-THESIS-2010-276} (2010).

\bibitem{ann2} 
 C. Glover, et al., \emph{Charged particle track reconstruction using artificial neural networks,} http://cds.cern.ch/record/400994/files/p639.pdf (1992).
 
 \bibitem{ann3} 
  B. Denby, et al., \emph{Fast triggering in high-energy physics experiments using hardware neural networks, } \emph{IEEE Trans. Neural Networks} {\bf 14.5} (2003) 1010.
  

\bibitem{scikit}  
  F. Pedregosa, \emph{Scikit-learn: Machine learning in Python,} \emph{J. Mach. Learn. Res. } {\bf 12} (2011) 2825.

\bibitem{mlp}scikit-learn MLPRegressor, \url{http://scikit-learn.org/stable/modules/generated/sklearn.neural_network.MLPRegressor.html}, 2018.

 
\bibitem{ku1966notes}
H. H. Ku, \emph{Notes on the use of propagation of error formulas,} \emph{J. Res. Natl. Bur. Stand.}  {\bf 70.4} (1966).


\bibitem{bevington2003data}
R. P. Bevington and D. K. Robinson, \emph{Data Reduction and Error Analysis,} {McGraw-Hill} (2003).

\bibitem{angres}  
D. Samuel, P. B. Onikeri, and L. P. Murgod, \emph{Angular resolution of stacked resistive plate chambers,} \emph{J. Cosmol. Astropart. Phys.} {\bf 2017.01} (2017) 058.

\bibitem{aging}   S. S. Bhide, et al., \emph{On aging problem of glass resistive plate chambers,} \emph{Nucl. Phys. B Proc. Suppl.} {\bf 158} (2006) 195.

\bibitem{aging2} A. Candela, et al., \emph{Ageing and recovering of glass RPC,}  \emph{Nucl.  Instrum. Methods Phys. Res., Sect. A} {\bf 533.1-2} (2004) 116.






\end{thebibliography}
\end{document}